\documentclass[dvipsnames,sort&compress,square,comma,authoryear]{aastex631}

\usepackage{hyperref}
\usepackage{wrapfig}
\usepackage{amsmath}
\usepackage{amssymb}
\usepackage{braket}
\usepackage{bm}
\usepackage{esdiff}

\def\vth{\vartheta}

\newcommand{\be}{\begin{equation}}
\newcommand{\ee}{\end{equation}}
\newcommand{\bea}{\begin{equation}\begin{aligned}}
\newcommand{\eea}{\end{aligned}\end{equation}}

\def\llangle{\left\langle}
\def\rrangle{\right\rangle}

\def\iso#1#2{\mbox{${}^{#2}{\rm #1}$}}
\def\be1#1{\iso{Be}{1#1}}
\def\n1#1{\iso{N}{1#1}}
\def\na2#1{\iso{Na}{2#1}}
\def\al2#1{\iso{Al}{2#1}}
\def\ar3#1{\iso{Ar}{3#1}}
\def\ca4#1{\iso{Ca}{4#1}}
\def\k4#1{\iso{K}{4#1}}
\def\mn5#1{\iso{Mn}{5#1}}
\def\ni5#1{\iso{Ni}{5#1}}
\def\ge6#1{\iso{Ge}{6#1}}
\def\fe6#1{\iso{Fe}{6#1}}
\def\rb8#1{\iso{Rb}{8#1}}
\def\nb9#1{\iso{Nb}{9#1}}
\def\zr9#1{\iso{Zr}{9#1}}
\def\ru9#1{\iso{Ru}{9#1}}
\def\mo9#1{\iso{Mo}{9#1}}
\def\tc9#1{\iso{Tc}{9#1}}
\def\pd10#1{\iso{Pd}{10#1}}
\def\cs13#1{\iso{Cs}{13#1}}
\def\sm14#1{\iso{Sm}{14#1}}
\def\gd15#1{\iso{Gd}{15#1}}
\def\dy15#1{\iso{Dy}{15#1}}
\def\hf18#1{\iso{Hf}{18#1}}
\def\gd15#1{\iso{Gd}{15#1}}
\def\pb20#1{\iso{Pb}{20#1}}
\def\bi20#1{\iso{Bi}{20#1}}
\def\u23#1{\iso{U}{23#1}}
\def\np23#1{\iso{Np}{23#1}}
\def\pu24#1{\iso{Pu}{24#1}}
\def\cm24#1{\iso{Cm}{24#1}}
\def\th23#1{\iso{Th}{23#1}}
\def\re18#1{\iso{Re}{18#1}}

\def\f1#1{\iso{F}{1#1}}
\def\b1#1{\iso{B}{1#1}}
\def\ba13#1{\iso{Ba}{13#1}}
\def\la13#1{\iso{La}{13#1}}
\def\ta18#1{\iso{Ta}{18#1}}

\begin{document}

\title{Collective neutrino oscillations and heavy-element nucleosynthesis in supernovae:\\ exploring potential effects of many-body neutrino correlations}

\correspondingauthor{Xilu Wang, Amol V. Patwardhan}
\email{wangxl@ihep.ac.cn, apatward@umn.edu}

\author[0000-0002-2999-0111]{A. Baha Balantekin}
\affil{Department of Physics,
University of Wisconsin, Madison, WI 53706, USA}
\affiliation{Network for Neutrinos, Nuclear Astrophysics, and Symmetries (N3AS), University of California, Berkeley, Berkeley, CA 94720, USA}

\author[0000-0002-2962-3055]{Michael J.~Cervia}
\affil{Department of Physics, George Washington University, Washington, DC 20052}
\affil{Department of Physics, University of Maryland, College Park, MD, USA 20742}

\author[0000-0002-2281-799X]{Amol V. Patwardhan}
\affil{SLAC National Accelerator Laboratory, 2575 Sand Hill Rd, Menlo Park, CA 94025, USA}
\affil{School of Physics and Astronomy, University of Minnesota, Minneapolis, MN 55455, USA}
\affiliation{Network for Neutrinos, Nuclear Astrophysics, and Symmetries (N3AS), University of California, Berkeley, Berkeley, CA 94720, USA}

\author[0000-0002-4729-8823]{Rebecca Surman}
\affiliation{Department of Physics and Astronomy, University of Notre Dame, Notre Dame, IN 46556, USA}
\affiliation{Network for Neutrinos, Nuclear Astrophysics, and Symmetries (N3AS), University of California, Berkeley, Berkeley, CA 94720, USA}

\author[0000-0002-5901-9879]{Xilu Wang}
\affil{Key Laboratory of Particle Astrophysics, Institute of High Energy Physics, Chinese Academy of Sciences,  Beijing, 100049, China}
\affiliation{Network for Neutrinos, Nuclear Astrophysics, and Symmetries (N3AS), University of California, Berkeley, Berkeley, CA 94720, USA}

\begin{abstract}
    In high-energy astrophysical processes involving compact objects, such as core-collapse supernovae or binary neutron star mergers, neutrinos play an important role in the synthesis of nuclides. Neutrinos in these environments can experience collective flavor oscillations driven by neutrino-neutrino interactions, including coherent forward scattering and incoherent (collisional) effects. Recently, there has been interest in exploring potential novel behaviors in collective oscillations of neutrinos by going beyond the one-particle effective or \lq mean-field\rq\ treatments. Here, we seek to explore implications of collective neutrino oscillations, in the mean-field treatment and beyond, for the nucleosynthesis yields in supernova environments with different astrophysical conditions and neutrino inputs. We find that collective oscillations can impact the operation of the $\nu p$-process and $r$-process nucleosynthesis in supernovae. The potential impact is particularly strong in high-entropy, proton-rich conditions, where we find that neutrino interactions can nudge an initial $\nu p$ process neutron rich, resulting in a unique combination of proton-rich low-mass nuclei as well as neutron-rich high-mass nuclei. We describe this neutrino-induced neutron capture process as the \lq\lq $\nu$i process\rq\rq. In addition, nontrivial quantum correlations among neutrinos, if present significantly, could lead to different nuclide yields compared to the corresponding mean-field oscillation treatments, by virtue of modifying the evolution of the relevant one-body neutrino observables.
\end{abstract}

\section{Introduction}

Compact object environments such as core-collapse supernovae and binary neutron star mergers can serve as useful laboratories for various kinds of microphysical phenomena. In particular, it is known that neutrinos play a key role in transport of energy, entropy, and lepton number in these environments [e.g.,~\cite{Mezzacappa:1999nt, Janka:2006fh, Zhang:2013lka, Kyutoku:2017voj, Richers:2017uvm, Burrows:2020qrp, Fuller:2022nbn, Foucart:2022bth}] and are therefore expected to exert a significant influence on the dynamical phenomena such as shock reheating or baryonic matter outflows. Additionally, some of the key processes governing neutrino transport in these environments, namely charged-current neutrino and antineutrino absorption and emission, are also responsible for regulating the free neutron and proton abundances (as well as possibly transmuting heavier nuclei) and are therefore key drivers of nucleosynthesis [e.g.,~\cite{Epstein1988, Woosley1990, McLaughlin+1996,Surman+2004, Frohlich2006c, Kajino2014, Martinez2017, Frohlich:2015spx, Roberts:2016igt, Langanke2019, 2019PrPNP.107..109K, Wang2023, Fischer:2023ebq, Xiong:2023uyb}]. Core-collapse supernovae, particularly during the neutrino-driven wind phase \citep{Duncan:1986ndw} in the aftermath of the explosion, have long been considered as a possible site for the synthesis of trans-iron elements [e.g., \cite{Woosley:1992ek, Meyer+1992, Woosley+1994, Takahashi+1994, Qian+1996, Hoffman+1997}]. This early work on nucleosynthesis in neutrino-driven winds had an emphasis on neutron-rich nuclide production via rapid neutron capture ($r$ process). More recently, these environments were also identified as candidate sites for proton-rich nucleosynthesis via the $\nu p$ process \citep{Frohlich2006,Pruet2006,Wanajo2006b}.

At the temperatures and densities that are characteristic of these environments, the capture rate of neutrinos on nucleons via the charged-current weak interaction is flavor-asymmetric. Aside from a small number of muon neutrinos in the high-energy tail of the distribution, only the electron flavor neutrinos and antineutrinos have charged-current interactions that are kinematically permitted\footnote{More recently, the role of muons in supernova and merger environments has received greater consideration [e.g.,~\cite{Bollig:2017lki, Guo:2020tgx, Fischer:2020vie, Loffredo:2022prq}]. This aspect would necessitate differentiating between $\mu$ and $\tau$ flavor neutrinos, adding yet another wrinkle to the potential impact of flavor oscillations on the dynamics and nucleosynthesis. Three-flavor oscillations have been shown to have potentially distinct signatures compared to the two-flavor case [e.g.,~\cite{Dasgupta:2007ws, Chakraborty:2019wxe, Chakraborty:2021oah, Capozzi:2020kge, Capozzi:2020syn, Capozzi:2022dtr, Shalgar:2021wlj, Siwach:2022xhx}].
Nevertheless, for simplicity, we do not consider any three-flavor effects here.}. As a result, it becomes important to consider the possible consequences of neutrino flavor evolution on the nucleosynthesis outcomes [e.g.,~\cite{Qian:1993dg, Yoshida2006, Yoshida2006b, Chakraborty:2009ej, Duan2011, Martinez2011, Martinez2014, Kajino:2012zz, Wu2014, Wu2015, Sasaki2017, Balantekin2018, Xiong2019, Ko:2019xxm, Xiong2020, George2020}]. Neutrinos in these environments are expected to exhibit various kinds of collective effects in flavor space, engendered by neutrino-neutrino ($\nu$-$\nu$) interactions [\cite{Pantaleone:1992eq,Pantaleone:1992xh,Samuel:1993,Sigl:1993ctk,Qian:1994wh}]. These effects can be driven either by coherent neutrino-neutrino forward scattering [e.g., \cite{Duan2009,Duan2010,Chakraborty+2016,Tamborra2020,Richers2022,Volpe:2023met} and references therein], as well as by collisional effects [e.g., \cite{Johns:2021qby,Johns:2022yqy,Lin:2022dek,Xiong:2022zqz,Padilla-Gay:2022wck}].

An ensemble of neutrinos where the flavor oscillations are driven by coherent $\nu$-$\nu$ interactions represents a quantum many-body problem in flavor space. Such a problem can in general be quite challenging, because the size of the problem (and the associated computational complexity) grows exponentially with the number of interacting particles in the ensemble. 
To overcome this limitation, the collective behavior of neutrinos is often analyzed in a one-particle effective description, wherein {each neutrino is represented by its own density matrix whose evolution is driven, at leading order, by interactions with all the other neutrinos---also represented as individual density matrices}. In this approach, the effect of multi-particle correlations or quantum entanglement between the individual neutrinos is absent at leading order in the Fermi coupling constant $G_F$.
This approximation reduces the scaling of the complexity to linear rather than exponential, making it possible to simulate much larger systems.

Exploring the potential presence of novel phenomena in the collective neutrino oscillation landscape beyond the one-particle effective description has been a topic of long-standing interest \citep{Bell:2003mg, Friedland:2003dv, Friedland:2003eh, Friedland:2006ke, McKellar:2009py, Balantekin:2006tg}. Recent studies, using simplified models of interacting neutrino ensembles, have sought to quantify these effects in terms of their impact on the relevant one-body neutrino observables, such as the net flavor survival/oscillation probabilities [e.g.,~\cite{Pehlivan:2011hp, Volpe:2013uxl, Pehlivan:2016lxx, Birol:2018qhx, Cervia:2019res, Rrapaj:2019pxz, Roggero:2021asb, Roggero:2021fyo, Patwardhan:2021rej, Roggero:2022hpy, Xiong2022, Martin:2021bri, Cervia:2022pro, Siwach:2022xhx, Martin:2023ljq, Martin:2023gbo}---or see \cite{Patwardhan:2022mxg, Balantekin:2023qvm, Volpe:2023met} and references therein]. Preliminary attempts have also been made to simulate the evolution of interacting neutrino systems using quantum computers \citep{Hall:2021rbv,Yeter-Aydeniz:2021olz,Illa:2022jqb,Amitrano:2022yyn,Illa:2022zgu,Siwach:2023wzy}.
Since these flavor survival/oscillation probabilities can directly couple to the neutron-to-proton ratio of the medium through which the neutrinos pass, any changes in their evolution arising due to the effects of many-body quantum correlations could have direct implications for the nucleosynthesis in these environments. Here, we investigate such potential effects {on nucleosynthesis} using a simple toy model for the interacting neutrino system. 

It has recently been argued by \cite{Shalgar:2023ooi} and \cite{Johns:2023ewj} that a more faithful assessment of beyond-the-mean-field effects in collective oscillations could require extending this framework to include additional physics, such as a finite neutrino wave-packet size and neutrino interactions beyond just forward and exchange scattering. These could constitute directions for future work in this field. Here, we work with the premise that our toy model is not intended to necessarily serve as an accurate representation of a stream of neutrinos propagating outward from a supernova core, because of the small number of interacting neutrinos considered in our system and a number of other simplifying assumptions (e.g., the interacting plane wave description, a trajectory-independent interaction strength, ignoring the matter potential in the neutrino Hamiltonian, etc.).  Our calculations are simply intended as an illustration of how \emph{any} potential changes in neutrino oscillation patterns, due to collective effects in the mean-field analysis and/or beyond, could have implications for nucleosynthesis in different physical regimes.

\section{Background}

We seek to examine an environment that represents the immediate aftermath of a core-collapse supernova explosion. In this situation, neutrinos are being emitted from the surface of a nascent protoneutron star (PNS), and some of these neutrinos interact with the surrounding medium and deposit their energy, heating the medium and driving outflows of baryonic matter---often called neutrino-driven winds~\citep{Duncan:1986ndw}. As the matter in these outflows expands and cools across a range of densities and temperatures, it can play host to a variety of different nucleosynthetic processes, depending on the physical conditions in the outflow and the neutrino emission characteristics [e.g., \cite{Woosley+1994, Qian+1996, Hoffman+1997, Balantekin:2004ug, Frohlich2006}; and many more]. Of particular relevance is the neutron-to-proton ratio (or, equivalently, the electron fraction $Y_e$) of the initial outflow, since it plays a key role in determining its eventual isotopic composition. This ratio is determined by a competition between electron neutrino and electron antineutrino capture rates.

\subsection{Neutrino capture rates} \label{sec:nucap}

Here, we shall describe how the charged-current neutrino and antineutrino capture rates are calculated in this simplified neutrino problem. These rates depend on the neutrino number fluxes across the energy spectrum and the corresponding cross-sections. For example, for $\nu_e$ capture on neutrons, one has
\begin{equation}
\lambda_{\nu_en}(\bm{r},t) = \int_{\bm p} \sigma_{\nu_en} (\bm{p})\, dn_{\nu_e}(\bm{r},t, \bm{p}) \,,
\label{eq:rate}
\end{equation}
where $\sigma_{\nu_en}$ is the capture cross-section and $dn_{\nu_e}(\bm{r},t,\bm{p})$ is the differential number density of electron neutrinos with momentum $\bm{p}$, at location $\bm{r}$ and time $t$. The limits of integration on the neutrino momentum directions in Eq.~\eqref{eq:rate} are dependent on the geometry of the neutrino source.  In the rest of the discussion, we shall assume that there is no explicit time-dependence in the problem, so that the geometry and the neutrino emission characteristics are fixed with respect to time\footnote{From the point of view of the flavor evolution calculations, the time-independent approximation can be justified on account of the neutrino crossing timescale across the supernova envelope being much shorter than the protoneutron star cooling timescale over which the neutrino emission characteristics change significantly. From the point of view of nucleosynthesis, a justification of this approximation requires the relevant expansion timescales of the neutrino driven wind to also be shorter than the PNS cooling timescale. This happens to be the case for the outflow trajectories considered here.}. Hence, the label $t$ shall be omitted henceforth. {The neutrino capture rates do still have an \textit{implicit} time dependence, encoded in the motion $r(t)$ of a fluid element as it travels outward from the proto-neutron star.}

\begin{figure}[htb]
    \centering
\includegraphics[width = 0.8\textwidth]{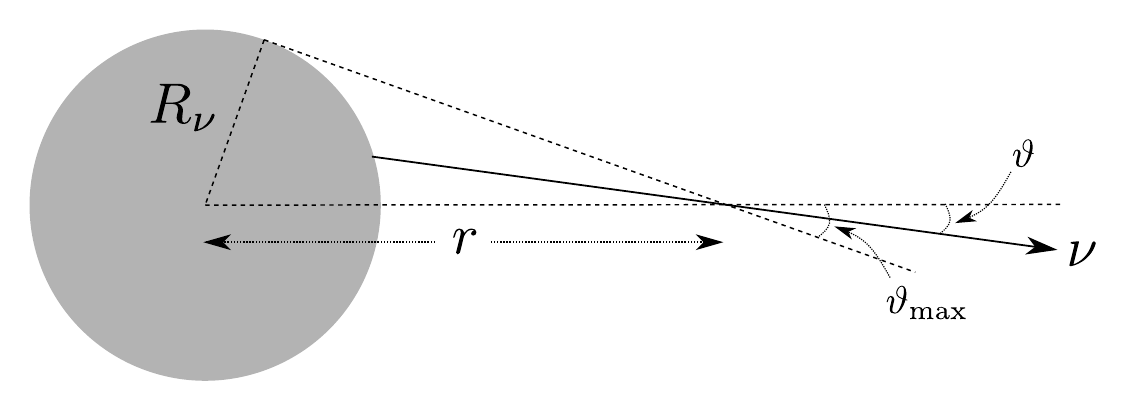}
    \caption{An illustration of the \lq\lq neutrino bulb\rq\rq --- a spherically symmetric neutrino source of radius $R_\nu$, with an emission that is assumed to be (semi-)isotropic from each point on the surface. This setup has been frequently adopted to study neutrino oscillations in supernovae. Neutrinos incident upon a location at a distance $r$ from the center of the sphere can be seen to be confined within a cone bounded by values of the polar angle $\vartheta \in [0,\vartheta_\text{max}]$.
    }
\label{fig:bulb}
\end{figure}

Here it is useful to define the quantity $dn_{\nu,\alpha}(\bm{r},\bm{p})$ as the differential number density of the subset of neutrinos at location $\bm{r}$ with momentum $\bm{p}$, which were emitted specifically with an initial flavor $\alpha$. In other words, $dn_{\nu,\alpha}(\bm{r},\bm{p})$ would represent the differential neutrino number density in flavor $\alpha$, \textit{in the absence of any flavor evolution}. One can then write $dn_{\nu_e}$ in terms of $dn_{\nu,\alpha}$ for each flavor and the corresponding flavor conversion probabilities $P_{\alpha e}$ as follows:
\begin{equation}
dn_{\nu_e}(\bm{r},\bm{p})= \sum_\alpha dn_{\nu,\alpha}(\bm{r},\bm{p})\, P_{\alpha e}(\bm{r},\bm{p}).
\end{equation}
In what follows, we shall assume a spherically symmetric neutrino source with a sharp decoupling surface at $r=R_\nu$ and isotropic emission in all outward directions from each point on the surface [i.e., the neutrino bulb geometry from \cite{Duan:2006an}].
Under this assumption, the differential number densities $dn_{\nu,\alpha}$ lose their dependence on the direction of the neutrino momenta and can instead be written simply in terms of the neutrino energy, as:
\begin{equation}
    dn_{\nu,\alpha}(r,E) = \frac{1}{2\pi^2 {(\hbar c)^3}} \frac{d\cos{\vth}\,d\phi}{4\pi} f_{\nu,\alpha}(E)\, E^2 dE, 
\end{equation}
where $f_{\nu,\alpha}(E)$ is the initial occupation number as a function of energy, and $\vth$ and $\phi$ are the polar and azimuthal angles, respectively, relative to the radial direction at distance $r$ from the center of the neutrino source. With these symmetry assumptions, $d\phi$ integrates to $2\pi$, and the limits of integration for $\cos\vartheta$ depend on $r$ (e.g., see Fig.~\ref{fig:bulb}). 

\subsection{Collective neutrino flavor oscillations}
\label{sec:mb-intro}

Flavor oscillations in dense neutrino streams may be substantially impacted by coherent $\nu$-$\nu$ forward scattering. This interaction causes the flavor evolution histories of neutrinos with different energies/momenta to become coupled to one another. Moreover, these interactions may also lead to quantum entanglement among neutrinos. In essence, an interacting neutrino system undergoing flavor evolution is a quantum many-body system described by a Hamiltonian with one- and two-body interactions, the latter of which are long-range in momentum space. {The \lq one-body\rq\ terms describe vacuum oscillations of each neutrino as well as neutrino coherent forward scattering with ordinary matter (electrons and nucleons)~\citep{Wolfenstein1978,Mikheyev1985}, whereas the two-body term represents $\nu$-$\nu$ interactions~\citep{Pantaleone:1992eq,Pantaleone:1992xh}. In regimes where $\nu$-$\nu$ interactions dominate, the neutrino-matter forward scattering term can often be \lq\lq rotated away\rq\rq\ with a suitable change-of-basis transformation, at the expense of an in-medium modification to the mixing angle and mass-squared splitting. For simplicity, we neglect this term in our calculation.} For a system of $N$ neutrinos with discrete momenta (and, therefore, discrete energies) quantized as plane waves in a box of volume $V$, this Hamiltonian in a two-flavor approximation can be written as:
\begin{equation} \label{eq:mbhamilt}
    H = \sum_{\bm p}  \omega_{\bm p} \, \vec{B} \cdot \vec{J}_{\bm p} + \sum_{\bm{p},\bm{q}} \mu_{\bm{pq}} \, \vec{J}_{\bm p} \cdot \vec{J}_{\bm q},
\end{equation}
where $\omega_{\bm p} \equiv \delta m^2/(2|\bm{p}|)$ are the vacuum oscillation frequencies of the neutrinos (with $\delta m^2$ being the mass-squared splitting). $\vec{J}_{\bm p}$ is a vector of operators forming an SU(2) algebra and representing the neutrino \lq\lq isospin\rq\rq\ in the flavor or mass basis [for details, see \cite{Balantekin:2006tg,Pehlivan:2011hp}]. Additionally, $\mu_{\bm{pq}} \equiv (\sqrt{2} G_F / V ) \, (1 - \cos{\theta_{\bm{pq}}})$ is the interaction strength between the neutrinos~\footnote{Some studies define the interaction strength as $\mu_{\bm{pq}} \equiv (\sqrt{2} G_F N / V)$, with the Hamiltonian terms then being proportional to $\mu_{\bm{pq}}/N$, thus achieving the same final outcome.}, which in general depends on the intersection angle $\theta_{\bm{pq}}$ between their momenta, and $\vec{B} = (0,0,-1)$ in the mass basis and $(\sin{2\theta},0,-\cos{2\theta})$ in the flavor basis, where $\theta$ is the neutrino mixing angle. 

We assume the neutrinos to be produced in an initially uncorrelated state at the PNS surface, and therefore their $N$-body wave function has a direct product form:
\begin{equation}
    \ket{\Psi(r = R_\nu)} = \bigotimes_{\bm p} \ket{\psi_{\bm p}}.
\end{equation}
Since the neutrinos are produced by weak interactions, one typically assumes that each $\ket{\psi_{\bm p}}$ is initially in a flavor state (i.e., $\ket{\nu_e}$ or $\ket{\nu_x}$ in the two-flavor approximation, where $\ket{\nu_x}$ is a superposition of $\ket{\nu_\mu}$ and $\ket{\nu_\tau}$). The wave function $\ket\Psi$ evolves in time according to the Schr\"odinger equation\footnote{Previously, we stated the assumption of steady-state neutrino emission, removing any explicit time-dependence from the problem. With that assumption, \lq\lq time\rq\rq\ $t$ in the Schr\"odinger equation can be treated as a proxy for location $\vec{r}$, with $t=0$ corresponding to $|\vec{r}|=R_\nu$. Moreover, in this spherically symmetric single-angle picture, we follow only radially directed trajectories for flavor evolution and nucleosynthesis.}:
\begin{equation} \label{eq:schrod}
    i\frac{\mathrm{d}}{\mathrm{d}t}\ket{\Psi} = H \ket{\Psi},
\end{equation}
where $H$ is the many-body neutrino Hamiltonian from Eq.~\eqref{eq:mbhamilt}. In general, the $\nu$--$\nu$ interaction term of the Hamiltonian can drive the evolution of the wave function away from a separable product state, resulting in development of nontrivial quantum entanglement between the constituent neutrinos. In this situation, a complete description of the many-body evolution as a function of time requires a Hilbert space that is $2^N$-dimensional, where $N$ is the number of neutrinos.

The flavor evolution of individual neutrinos in the ensemble can be quantified in terms of the expectation values of certain one-body operators within each individual neutrino subspace---for instance, the components of the isospin $\vec{J}_{\bm q}$ in the flavor or mass basis. One can define the neutrino \lq\lq Polarization vectors\rq\rq\ $\vec{P}_{\bm q} = 2 \, \langle \vec{J}_{\bm q} \rangle$, which contain information about the flavor composition of the neutrino, with $P_z$ denoting the net lepton number in the flavor or mass basis and the $P_x$ and $P_y$ components representing the coherence between different flavor or mass eigenstates. The relation between the Polarization vector components in the flavor ($f$) and mass ($m$) bases is given by:
\begin{align}
    P_z^{(f)} &= \cos(2\theta) \, P_z^{(m)} + \sin(2\theta) \, P_x^{(m)}, \\
    P_x^{(f)} &= \cos(2\theta) \, P_x^{(m)} - \sin(2\theta) \, P_z^{(m)}, \\
    P_y^{(f)} &= P_y^{(m)}, 
\end{align}
where $\theta$ is the neutrino mixing angle. The electron flavor survival probability, which is required for calculating the (anti-)neutrino capture rates, can then be obtained using $P_{\alpha e} = (1 + \text{sign}(\omega) \, P_z^{(f)})/2$. Here, $\text{sign}(\omega) = +1$ for neutrinos and $-1$ for antineutrinos.

\subsection{The mean-field approximation}
\label{sec:mf-intro}

The exponential scaling of the complexity of this many-body system as a function of particle number makes this problem numerically intractable even for a relatively small number ($\sim$ a few tens) of neutrinos in the most general case. To circumvent this difficulty, a frequently adopted approach involves neglecting any nontrivial quantum correlations among the neutrinos, thereby {enforcing that the neutrino wave function (initially in a product state) forever remains in a product state}\footnote{Quantum Kinetic treatments [e.g.,~\cite{Sigl:1993ctk,Vlasenko:2013fja}] that include both coherent flavor evolution and incoherent collisional effects do implicitly capture some effects of multi-neutrino quantum correlations up to $\mathcal{O}(G_F^2)$. Here, \lq\lq mean-field\rq\rq\ refers to just the coherent part of the quantum kinetic description.}. Effectively, this simplification amounts to replacing operator products in the Hamiltonian with one-body terms proportional to an operator times an expectation value. 
One can then reduce the Hamiltonian to an effective one-body operator form [\cite{Balantekin:2018mpq}]:
\begin{equation} \label{eq:obhamilt}
    H = \sum_{\bm p}  \omega_{\bm p} \, \vec{B} \cdot \vec{J}_{\bm p} + \sum_{\bm{p},\bm{q}} \mu_{\bm{pq}} \, \left[ \vec{J}_{\bm p} \cdot \langle \vec{J}_{\bm q} \rangle + \langle \vec{J}_{\bm p} \rangle \cdot \vec{J}_{\bm q} - \langle \vec{J}_{\bm p} \rangle \cdot \langle \vec{J}_{\bm q} \rangle \right].
\end{equation}
Ehrenfest's theorem then dictates that the expectation values of the isospin operators, $\langle \vec{J}_{\bm q} \rangle \equiv \bra{\Psi} \vec{J}_{\bm q} \ket{\Psi}$ evolve according to:
\begin{equation}
    \diff{\langle \vec{J}_{\bm q} \rangle}{t} = -i \llangle \left[\vec{J}_{\bm q},H \right] \rrangle.
\end{equation}
Using the SU(2) commutation relations for the isospin operators, together with the mean-field condition from Eq.~\eqref{eq:obhamilt}, one obtains the mean-field evolution equations for the neutrino polarization vectors, $\vec{P}_{\bm q}$: 
\begin{equation} \label{eq:mfevol}
    \diff{\vec{P}_{\bm q}}{t} = \omega_{\bm q} \, \vec{B} \times \vec{P}_{\bm q} + \sum_{\bm p} \mu_{\bm{pq}} \, \vec{P}_{\bm p} \times \vec{P}_{\bm q}.
\end{equation}

Certain practical differences between the many-body and the mean-field treatments can be quantified by comparison of their predictions for the evolution of Polarization vectors. 
In the many-body treatment, one can use Eq.~\eqref{eq:schrod}, together with the Hamiltonian from Eq.~\eqref{eq:mbhamilt}, to evolve the many-body wave function $\ket\Psi$ and then compute the polarization vectors $\vec{P}_{\bm q} = 2 \, \langle \vec{J}_{\bm q} \rangle$ at each time step. 
These results can then be compared to the Polarization vectors of the corresponding mean-field solution, obtained by integrating Eq.~\eqref{eq:mfevol}.

\section{Implementation}

Often, for the sake of simplicity, studies of collective neutrino oscillations employ the \textit{single-angle approximation}, wherein the flavor evolution of neutrinos is taken to be trajectory-independent. The interaction strength $\mu_{\bm{pq}}$ can then be replaced with a single suitable angle-averaged interaction parameter $\mu$, which depends on on the distance from the source, i.e., $\mu = \mu(r)$. For the \lq\lq neutrino bulb\rq\rq\ setup described in Sec.~\ref{sec:nucap} and depicted in Fig.~\ref{fig:bulb}, the interaction strength varies with radius as [\cite{Duan:2006an}]: 
\begin{equation} \label{eq:mur}
    \mu(r) = \mu(R_\nu) \left[1 - \sqrt{1-\left(\frac{R_\nu}{r}\right)^2}\right]^2,
\end{equation}
where the normalization factor $\mu(R_\nu)$ can be determined from the neutrino emission characteristics (i.e., luminosity and average energy) at the emission surface ($r = R_\nu$). In our calculations of the neutrino flavor evolution, we treat $\mu(R_\nu)$ simply as a numerical parameter. {However, in order to be able to use the flavor evolution results to consistently compute the charged-current neutrino capture rates in physical units, we impose that
the luminosities $L_{\nu,\alpha}$ of neutrinos with initial flavor $\alpha$ each have to satisfy the relation:
\begin{equation} \label{eq:murnu}
    \mu(R_\nu) = \frac{\sqrt2 \, G_F \, L_{\nu,\alpha}}{4\pi R_\nu^2 \, \langle E_{\nu,\alpha} \rangle  N_{\nu,\alpha}} 
\end{equation}
where $N_{\nu,\alpha}$ is the total number of neutrinos in flavor $\alpha$ in the initial condition of the system, and $\braket{E_{\nu,\alpha}}$ is the corresponding initial average energy\footnote{At a quick glance, the right-hand side in the above equation appears to be $\alpha$-dependent, yet the left-hand side is not. This is easily resolved by recognizing that the ratio $L_{\nu,\alpha}/\langle E_{\nu,\alpha} \rangle$ is proportional to the neutrino number $N_{\nu,\alpha}$ in initial flavor $\alpha$.}. 
For the numerical implementation, we take a neutrino spectrum comprised of $N$ discrete neutrino energy modes, with exactly one (anti-)neutrino occupying each mode. We take half of the modes to be occupied by neutrinos and the other half by antineutrinos. In that case, Eq.~\eqref{eq:rate} is replaced by its discretized version:
\begin{equation}
    \lambda_{\nu_en}(r) = \sum_E \sigma_{\nu_en}(E) \, n_{\nu_e}(r,E) \,,
\label{eq:discreterate}
\end{equation}
where $n_{\nu_e}(r,E) = \sum_\alpha n_{\nu,\alpha}(r,E) P_{\alpha e}(r,E)$. Since the neutrino calculations in the many-body case are computationally intensive due to the exponential scaling of the Hilbert space, we use for the purposes of illustration only a small number of neutrino modes ($N = 4$ for most calculations, and a couple of cases with $N = 8$ for comparison). Each mode is assigned a vacuum oscillation frequency $\omega_q$ that corresponds to physically reasonable choices of neutrino energy, in the 10--20\,MeV range.

\begin{table}[!htb]
\centering
\caption{
    \label{parameters}
    Summary of parameters used in our calculations of neutrino oscillations. Coupling among neutrinos $\mu(R_\nu)$ is related to the physical parameters of the neutrino emission through Eq.~\eqref{eq:murnu}. The reference scale $\omega_0$ is defined in terms of the mass-squared splitting $\delta m^2 = 2.44\times10^{-3}$\,{eV${}^2$}, and a reference energy $E_0 = 10$ MeV.
}
\begin{tabular}{|l|r|r|}
\hline
{Parameter} 	&	\multicolumn{2}{c|}{{Value}} \\ \cline{2-3}
 {} & {Number} & {Unit}	\\ 
 \hline
$\omega_0$          &	$1.22\times10^{-16}$    &	MeV	        \\
\hline
$\sin^2(2\theta)$	&	$0.10$		            &				\\
\hline
$R_\nu$		        &	$15$	                &	km	        \\
\hline
$\mu(R_\nu)$        &   $6.954 \times10^{5}$     &  $ \omega_0$  \\
\hline
\end{tabular}
\end{table}

We start our flavor calculations at a radius $r_i$, where $\mu(r_i) = 100\,\omega_0$ for the $N=4$ calculations, or $50\,\omega_0$ for the $N=8$ calculations, so that $\mu(r_i)N$, which is proportional to the physical neutrino number density, is the same in both cases. $\omega_0$ is a certain reference frequency, with an associated physical scale that is given in Table~\ref{parameters}.
The total number density at radius $r$, of neutrinos emitted initially in flavor $\alpha$, can likewise be written as:
\begin{equation} \label{eq:ntotnorm}
    n_{\nu,\alpha}(r) = \frac{L_{\nu,\alpha}}{2\pi R_\nu^2 \, \langle E_{\nu,\alpha} \rangle} \left[1 - \sqrt{1-\left(\frac{R_\nu}{r}\right)^2}\right] \equiv n_{\nu,\alpha}(R_\nu) \left[1 - \sqrt{1-\left(\frac{R_\nu}{r}\right)^2}\right].
\end{equation}
Table~\ref{parameters} has a summary of the relevant neutrino parameters and physical scales used in our calculations. The numerical computation of the neutrino flavor evolution was performed using a fourth-order Runge-Kutta method (RK4) with an adaptive time-step that is varied with the interaction strength $\mu$, in both the many-body and the mean-field configurations. 
Finally, the output of this calculation, in the form of the integrated neutrino and antineutrino capture rates as a function of radius, was used as an input to the nuclear reaction network. 

\section{Models} 
\label{sec:model}

\subsection{Neutrino calculations}
\label{sec:nucalc}

Neutrinos play a vital role in supernova neutrino-driven wind nucleosynthesis processes through capture reactions on free nucleons:  
\begin{eqnarray} \label{eq:nucap}
\nu_e + n \rightleftharpoons p + e^{-},\\
\bar{\nu}_e + p \rightleftharpoons n + e^{+}.
\end{eqnarray}
These capture reactions shape the neutron-to-proton ratio ($n_{n}/n_{p}$) and the electron fraction $Y_{e}=1/(1+n_{n}/n_{p})$ in the initial phase of the outflow (at high temperature and density), which determine what type of nucleosynthesis is possible in the ejecta, and influence the availability of free nucleons to capture as element synthesis proceeds. 
We start each nucleosynthesis calculation from $T\sim10$ GK, when both electron and positron capture rates are very small compared to neutrino capture rates and the ejecta are in a weak equilibrium stage. At this stage, the electron fraction $Y_e$ can be approximated as 
\begin{equation}
Y_e \approx \frac{1}{1+\lambda_{\bar{\nu}_e}/\lambda_{\nu_e}}.
\label{eq:ye}
\end{equation}
In this work, we have investigated the nucleosynthesis outcomes for different choices of initial neutrino energies (or equivalently, oscillation frequencies $\omega_q$), {roughly based on supernova simulations (e.g., \cite{Keil+2003,Mirizzi+2016,Horiuchi+2018}).} Broadly, these choices are classified as \lq symmetric\rq\ (sym) or \lq asymmetric\rq\ (asym), based on whether the initial $\bar\nu_e$ average energies are equal to or greater than the initial $\nu_e$ average energies. For the calculations with $N=4$ neutrino modes, we assign one mode each to the $\nu_e$, $\bar\nu_e$, $\nu_x$, and $\bar\nu_x$ species, respectively ($N_{\nu,\alpha} = 1$ for each $\alpha$). For the $N=8$ case, we assign two modes to each of the species ($N_{\nu,\alpha} = 2$). Table~\ref{tab:neutrinomodels} summarizes the parameters for each of these initial neutrino distribution choices: the energies $E_{\nu,\alpha}$ and the corresponding luminosities $L_{\nu,\alpha}$ such that Eq.~\eqref{eq:murnu} is satisfied, given the values of $R_\nu$ and $\mu(R_\nu)$ from Table~\ref{parameters}. The resulting initial electron fractions to be used for the heavy-element nucleosynthesis simulations are also included in the final column.

\begin{table}[!htb]
    \centering
    \caption{\it 
        Initial conditions for neutrino energy distributions and electron fraction. 
        We initiate the oscillations at  $r_i\simeq 100$\,km, where $\mu_{i}=100\,\omega_0$ ($50\,\omega_0$ for the $4 \nu + 4 \bar\nu$ cases).
\label{tab:neutrinomodels}
    }
    \scalebox{1}{
    \hspace{-2.6cm}
    \begin{tabular}{|c|cccc|ccc|c|}
      \hline
      Label & $E_{\nu,{e}}$  & $E_{\bar\nu,{e}}$   & $E_{\nu,{x}}$  & $E_{\bar\nu,{x}}$ & $L_{\nu,{e}}$  & $L_{\bar\nu,{e}}$   & $L_{\nu,{x}}$ & $Y_e$   \\
      & [MeV] & [MeV]  & [MeV] & [MeV] & [erg/s] & [erg/s]  & [erg/s] & $(1+\lambda_{\bar\nu_{e}}/\lambda_{\nu_{e}})^{-1}$  \\
     \hline
    sym & 10 & 10 & 20 & 20 & $9.091 \times 10^{51}$ & $9.091 \times 10^{51}$ & $1.818 \times 10^{52}$ & 0.634  \\
    \hline
    asym2 & 10 & 12.5 & 20 & 20 & $9.091 \times 10^{51}$ & $1.136 \times 10^{52}$ & $1.818 \times 10^{52}$ & 0.504  \\
    \hline
    asym2.1 & 10 & 13 & 20 & 20 & $9.091 \times 10^{51}$ & $1.182 \times 10^{52}$ & $1.818 \times 10^{52}$ & 0.482 \\
    \hline
    asym3 & 10 & 14.28 & 20 & 20 & $9.091 \times 10^{51}$ & $1.298 \times 10^{52}$ & $1.818 \times 10^{52}$ & 0.427 \\
    \hline
    asym4 & 10 & 16 & 20 & 20 & $9.091 \times 10^{51}$ & $1.455 \times 10^{52}$ & $1.818 \times 10^{52}$ & 0.366 \\
    \hline
    \hline
    sym-4nu &  10, 11.11 &  10, 11.11 &  16.67, 20 & 16.67, 20 & $9.591 \times 10^{51}$ & $9.591 \times 10^{51}$ & $1.667 \times 10^{52}$ & 0.634 \\
    \hline
    asym2.1-4nu & 10, 11.11 & 12.8, 14.3 & 16.67, 20 & 16.67, 20 & $9.591 \times 10^{51}$ & $1.232 \times 10^{52}$ & $1.667 \times 10^{52}$ & 0.482 \\
    \hline
    \end{tabular}}
\end{table}

\begin{figure}[!htb]
	\centering
	\vspace{-5mm}
\includegraphics[width=8cm]{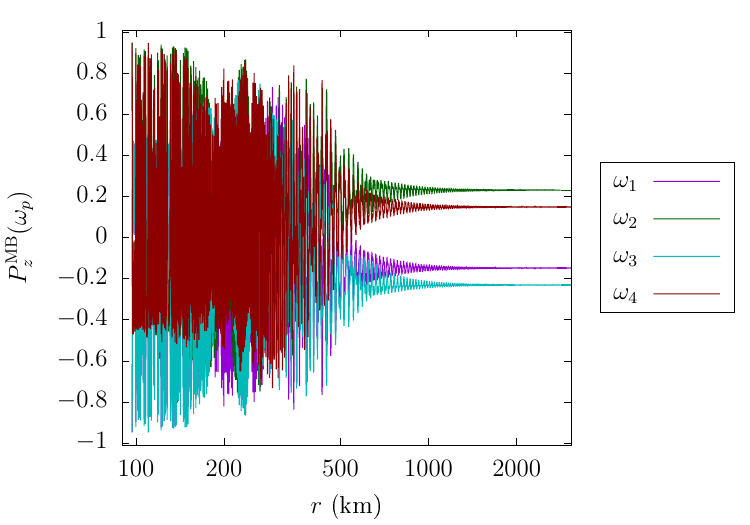}
\includegraphics[width=8cm]{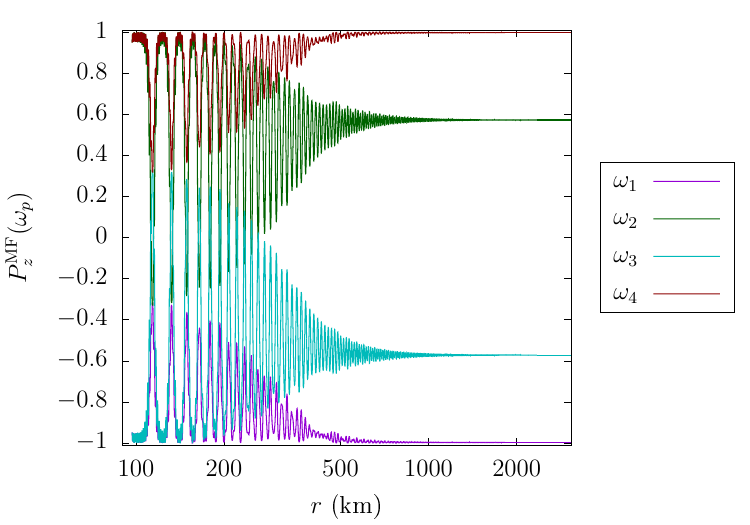}\\
\includegraphics[width=8cm]{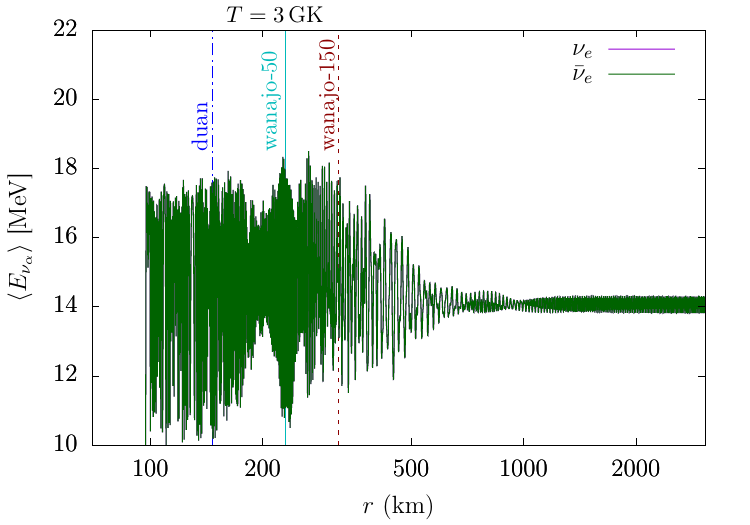}
\includegraphics[width=8cm]{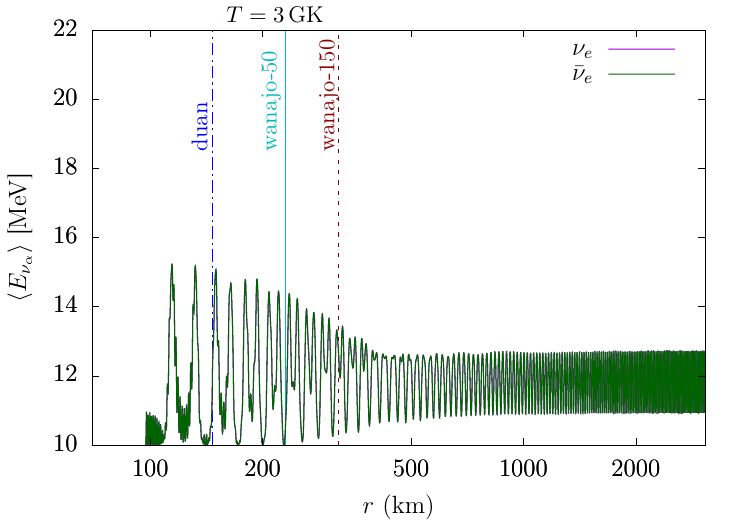}\\
\includegraphics[width=8cm]{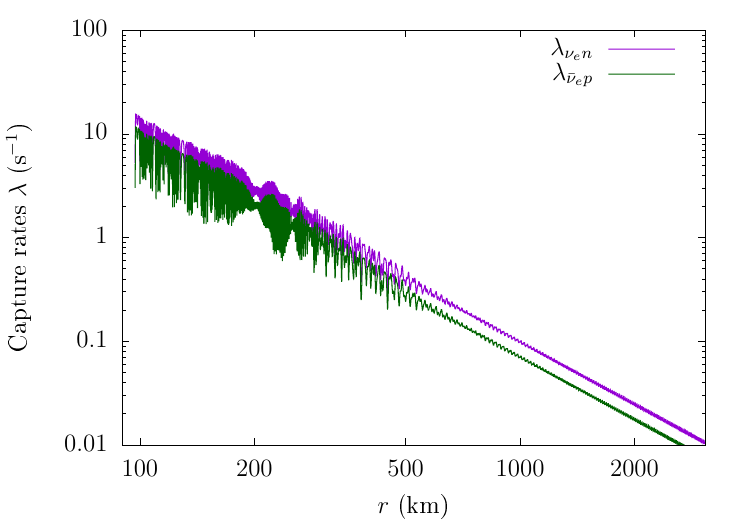}
\includegraphics[width=8cm]{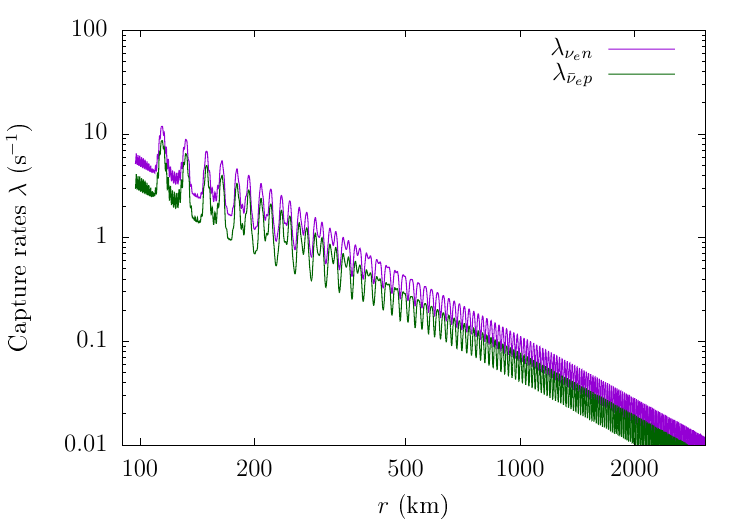}\\
	\caption{Evolution of mass-basis $P_z(\omega)$ (\textit{top panels}), {the average $\nu_e$ and $\bar\nu_e$ energies (\textit{middle panels})}, and the $\nu_e$ and $\bar\nu_e$ capture rates (\textit{bottom panels}) as functions of $r$ for a four-neutrino calculation (2 $\nu$ + 2 $\bar\nu$) in the \lq\lq sym\rq\rq\ configuration. The left panels show the results for a many-body calculation, whereas the right panels show the results for the corresponding mean-field calculation. {The locations at which three example matter trajectories (summarized in Table 3) reach a temperature of $T=3$\,GK---the approximate onset of a $\nu p$ process---are shown by vertical lines in the middle panels.} Note that, in the symmetric configuration, the average energies for $\nu_e$ and $\bar\nu_e$ are the same, but the corresponding capture rates are different due to the difference in $Q$-values for $\nu_e n$ and $\bar\nu_e p$ reactions. 
	}
	\label{fig:caprates0}
\end{figure}

To test the potential effects of neutrino interactions on the nucleosynthesis results, for each set of calculations, we consider four different neutrino treatments: no neutrinos involved for $T<10$ GK (nn), no neutrino oscillations (nosc), mean-field neutrino oscillations (mf) described in Sec.~\ref{sec:mf-intro}, and many-body neutrino oscillations (mb) described in Sec.~\ref{sec:mb-intro}. 
The default neutrino number used for our calculations is 2 $\nu$ + 2 $\bar\nu$; we also tested a symmetric (sym) and an asymmetric (asym2) scenario with an increased neutrino number: 4 $\nu$ + 4 $\bar\nu$. 
{Here we adopted the normal mass ordering for the neutrino flavor evolution as default (i.e., $\omega_{\bm p} > 0$ for neutrinos and $< 0 $ for anti-neutrinos); the effect of inverted mass hierarchy scenario is also examined, as discussed in Section~\ref{sec:sym-vary}.}

Figure~\ref{fig:caprates0} shows an example comparison between the mb and mf calculations for the flavor evolution of a 2 $\nu$ + 2 $\bar\nu$ neutrino system in the “sym” configuration. 
In particular, we show the evolution (i.e., with distance $r$) of neutrino polarization vector component $P_z(\omega)$ in the mass basis, evolution of the average $\nu_e$ and $\bar\nu_e$ energies, as well as of the $\nu_e$ and $\bar{\nu}_e$ capture rates. 
The vacuum oscillation frequencies of the neutrinos, calculated using their average energy through $\omega_j = \delta m^2/2E_j$, are $\omega_1$, $\omega_2$, $\omega_3$, and $\omega_4$ for neutrinos initially in the $\bar{\nu}_e$, $\bar{\nu}_x$, $\nu_x$, and $\nu_e$ states, respectively. We can see that the onset of high-amplitude oscillations occurs earlier for the mb case, and the two cases have distinct evolution with distance from the PNS, which we anticipate will influence the ultimate nucleosynthesis yields.

\subsection{Nucleosynthesis calculations}

In this work we adopt the nuclear reaction network code Portable Routines for Integrated nucleoSynthesis Modeling (PRISM)~\citep{Mumpower2018, Sprouse2020}. The four neutrino treatments summarized in Sec.~\ref{sec:nucalc}} (nn, nosc, mf, mb) are implemented into the nucleosynthesis calculations in the form of external $\nu_{e}+n$ and $\bar{\nu}_{e}+p$ capture rates, for each of the neutrino energy configurations from Table~\ref{tab:neutrinomodels}. At $T\sim10$\,GK, the nuclear composition is determined by Nuclear Statistical Equilibrium (NSE). Here we obtain the initial composition at $\sim10$\,GK through an NSE calculation with $Y_e$ obtained from Eq.~\eqref{eq:ye} and the density taken from the chosen matter trajectory. {We dynamically evolve the composition and the electron fraction from $T\sim 10$\,GK onwards. This naturally incorporates a possible `alpha effect'---an increase in the number of alpha particles and, subsequently, seeds and corresponding decrease in free nucleons available for capture---due to neutrino interactions on free nucleons during alpha particle formation \citep{Fuller+1995,Meyer1998}. } For the network calculation, we use REACLIB reaction rates \citep{Cyburt_2010} along with NUBASE $\beta$-decay properties \citep{Kondev_2021}. {The available REACLIB data is supplemented with neutron capture rates, $\beta$-decay rates, and fission properties in the actinide region based on the FRDM/FRLDM model \citep{Mumpower2018}, though we find the nucleosynthesis calculations in this work largely do not extend into this region of the nuclear chart.}

\begin{table}[!htb]
    \centering
    \caption{\it Supernova neutrino driven wind models adopted for our nucleosynthesis calculations.\\
    a. time elapsed from $T=6$\,GK to $T=3$\,GK.\\
    b. time elapsed from $T=3$\,GK to $T=1.5$\,GK.
\label{tab:nucleosynthesismodels}}
    \scalebox{0.95}{ \hspace{-2.6cm}
    \begin{tabular}{|c|c|c|c|c|}
      \hline
      Simulation Models & {Entropy per nucleon}  & \multicolumn{2}{|c|}{ Dynamical timescale}  & Initial radial distance  \\
      & {[s/k]} &  $\tau_1$\,[ms]$^a$ & $\tau_2$\,[ms]$^b$ &  $r_0$ [km]  \\
    \hline
     parameterization of & 50 & 17.5 & 152  & 61.58  \\
    -Wanajo2011 & 100 & 17.5 & 344  & 77.44 \\
     \citep{Wanajo2011} & 150 & 17.5 & 500  & 86.41  \\
    \hline
    Duan2011 \citep{Duan2011} & 200 & 12.4 & 17.9 & 46.67 \\
    \hline\hline
    \end{tabular}}
\end{table}

To test the nucleosynthetic outcomes of both the symmetric and asymmetric neutrino calculations, here we adopt two types of parameterized supernova neutrino-driven wind trajectories: trajectories parameterized as per the \lq\lq standard model\rq\rq\  in  \citet{Wanajo2011} (Wanajo2011) with a range of entropies, and a high-entropy trajectory from \citet{Duan2011} (Duan2011). The details of these trajectories are summarized in Table \ref{tab:nucleosynthesismodels}. We also tested the shocked trajectories from \citet{arc07}; the resulting abundance patterns lie between those for the Wanajo2011 trajectory with entropy per nucleon in units of the Boltzmann constant $s/k=50$, and those for the Duan2011 trajectory, so for brevity these results do not appear in this work.

The Wanajo2011 trajectories adopted the wind solution for the \lq\lq standard model\rq\rq\ in \citet{Wanajo2011} with $M_{\rm ns}=1.4\,M_\odot$ and $L_{\nu}=1\times10^{52}$\,erg/s, followed by a subsonic expansion after wind termination at 300 km using a standard parameterization with the time evolution of temperature $T$ and density $\rho$ as follows: $\rho(t)\propto t^{-2}$, and $T(t)\propto t^{-2/3}$. We assume an adiabatic expansion throughout the evolution, with constant values for the entropy per baryon set to $s/k=50$, 100, and 150. Higher (lower) values of initial entropy result in lower (higher) mass fractions of seed nuclei produced, with more (fewer) free nucleons remaining for capture. The Duan2011 trajectory is more extreme, with faster expansion and a higher entropy of $s/k\sim 200$. The density initially falls exponentially, $\rho(t)\propto e^{-t/3\tau}$, with $\tau\sim12$\,ms for $3\,\text{GK}\le T\le6\,\text{GK}$ and $\tau\sim18$\,ms for $1.5\,\text{GK}\le T\le 3\,\text{GK}$, before switching to a power law $\rho(t)\propto t^{-2}$ when the temperature is roughly 1\,GK. {This $\sim t^{-2}$ drop-off, rather than the much faster $\rho \propto t^{-3}$ that one expects in the event of free expansion, can be attributed to the fact that the neutrino-driven wind is expected to encounter the outer layers of the supernova ejecta/envelope as it emerges from the inner \lq\lq hot-bubble\rq\rq\ region, slowing down its expansion.} The entropy of this trajectory is somewhat outside of those expected from modern supernova simulations. However, this combination of fast expansion and high entropy maximizes the ratio of free nucleons to seeds, which is ideal for emphasizing the nucleosynthetic impact of the differences in neutrino treatments.

\begin{figure}[!htb]
	\centering
	\vspace{-5mm}
\includegraphics[width=8.5cm]{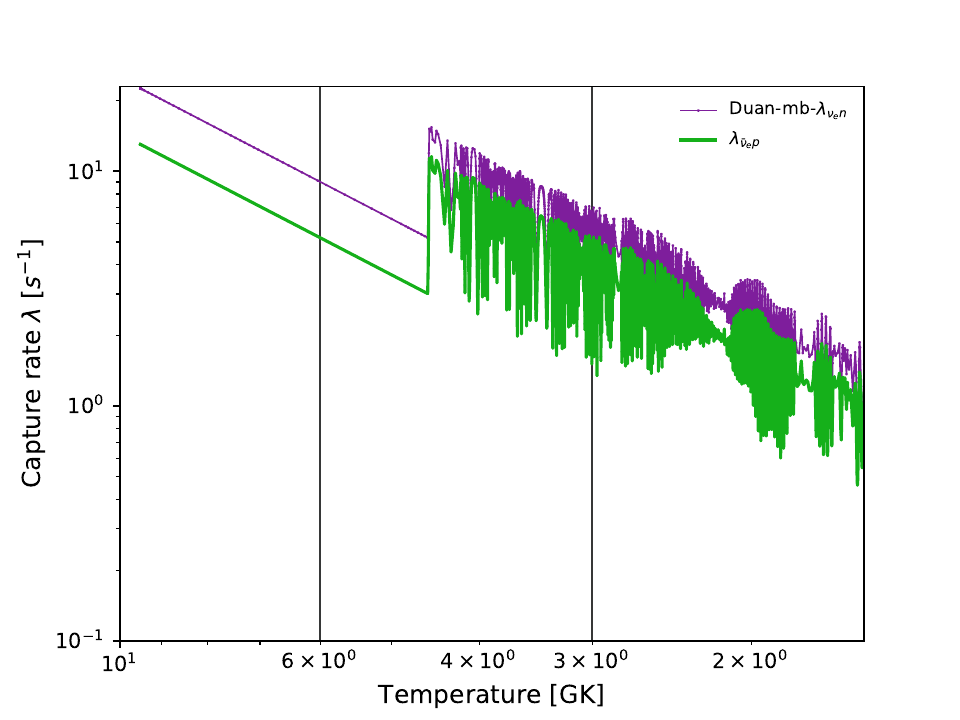}
\includegraphics[width=8.5cm]{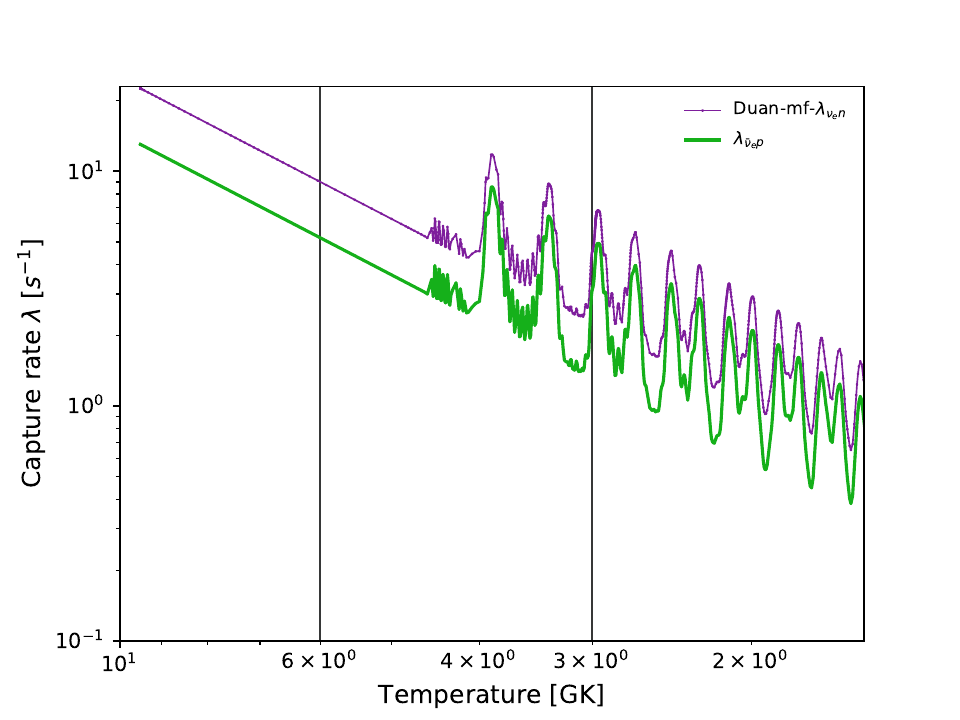}\\
\includegraphics[width=8.5cm]{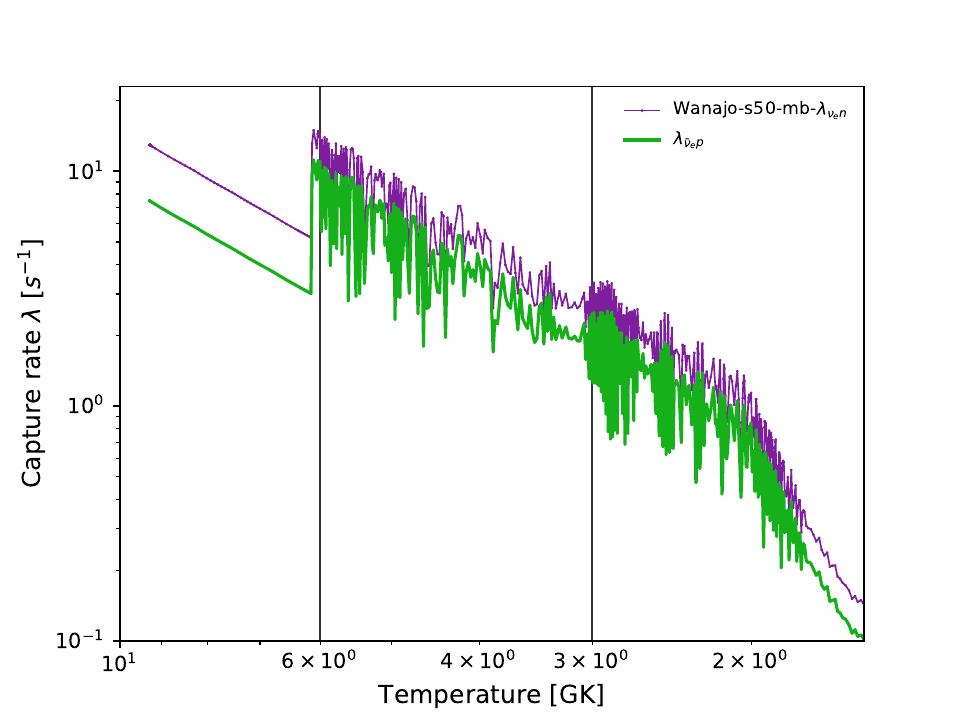}
\includegraphics[width=8.5cm]{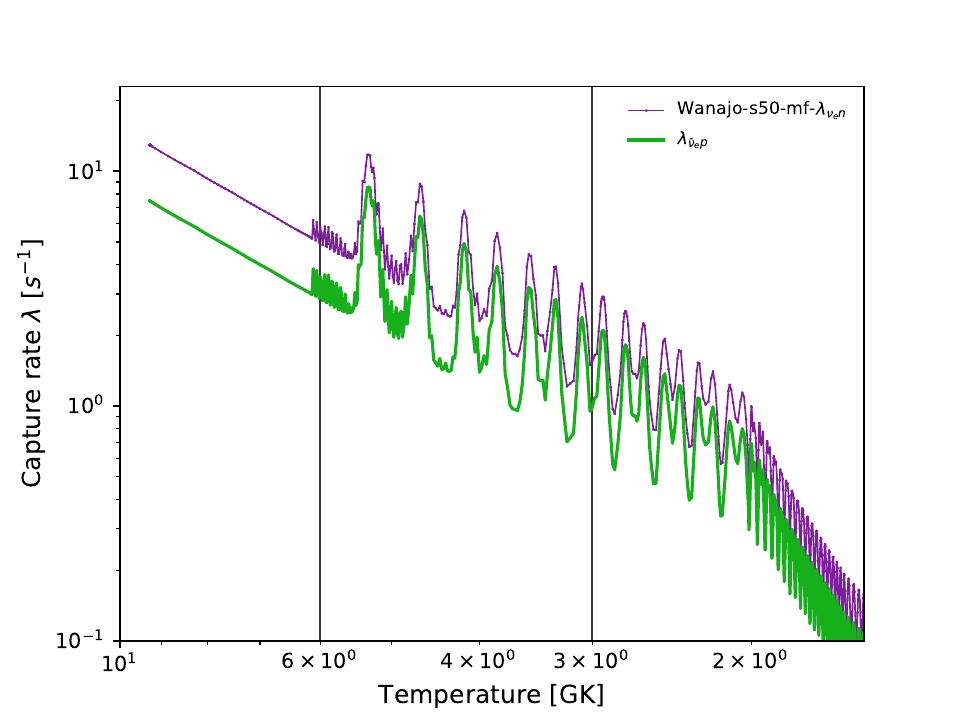}\\
	\caption{Neutrino and antineutrino capture rates for the symmetric initial conditions (left panels show a many-body treatment, while right panels show a mean-field treatment), shown versus temperature along the matter trajectory for Duan2011 (\textit{Top}) and Wanajo2011 (\textit{Bottom}). The vertical lines (including the right-sided plot boundary) indicate the temperature at 6 GK, 3 GK, and 1.5 GK, corresponding to the dynamical timescales estimated in Table~\ref{tab:nucleosynthesismodels}.
	}
	\label{fig:neutrino_duan_wanajo}
\end{figure}

Figure~\ref{fig:neutrino_duan_wanajo} illustrates the evolution of the neutrino capture rates as a function of the temperature along the matter trajectory for the Duan2011 and  Wanajo2011 $s/k=50$ cases, respectively. The neutrino fluxes are taken from the symmetric (sym) case with many-body (mb) and mean-field (mf) oscillation treatments. In all cases, the onset of oscillations results in a steep rise to the capture rates, which occurs within the temperature window relevant for nucleosynthesis.  The degree and duration of the rise are expected to shape the formation of seed nuclei and the subsequent nucleosynthesis. Note that, in each of our calculations, we assume steady-state neutrino emission from the source, resulting in an over-estimation of about 20\% in the integrated $\bar\nu_e$ capture rates for the Wanajo trajectories, and about 2\% for the Duan2011 trajectory, when compared with corresponding calculations that used exponentially decaying neutrino luminosities with a time constant of 3\,s. In each case, this approximation does not qualitatively affect any of the results and conclusions.

\section{Results} 
\label{sec:result}

The CCSN neutrino-driven wind was first believed to be a favored site for an $r$ process as a few early supernova models in the early 1990s predicted sufficiently high entropy and neutron-rich conditions \citep{Meyer+1992,Woosley+1994,Takahashi+1994}, although there were early indications that such conditions were difficult to attain \citep{Fuller+1995,McLaughlin+1996,Qian+1996,Hoffman+1997,Otsuki+2000,Thompson+2001,Terasawa+2002}. Recently, the modern simulations of CCSN neutrino-driven winds with neutrino transport find that the neutron-rich and high-entropy conditions needed for a strong $r$ process are unlikely~\citep{Arcones2007, Arcones2011, Fischer2010,Hudepohl2010, Roberts2010, Arcones2011b, Arcones2013, Bliss2018, Akram2020}. Still, an $r$ process, especially the weak $r$ process (producing nuclei up to $A\sim125$), might be possible in this environment, with the ultimate extent of the nucleosynthesis in this site sensitively dependent on neutrino physics~\citep{Fuller+1995,Balantekin:2004ug,Duan2011,Xiong2020}. Additionally, these modern simulations discovered that most or all of the neutrino-driven wind ejecta tend to be proton-rich with low to moderate entropies  \citep{Arcones2007,Arnould2003, Hudepohl2010, Janka2016, Fischer2020a}, introducing the possibility of neutrino-enhanced, proton-rich nucleosynthesis -- the $\nu p$ process \citep{Frohlich2006, Pruet2006, Wanajo2006b}. Here we consider both neutron-rich and proton-rich nucleosynthetic scenarios.

\subsection{Heavy-element nucleosynthesis with symmetric neutrino energies}
\label{sec:sym}

Symmetric neutrino energies and luminosities between the $\nu_e$ and $\bar\nu_e$ species will produce charged-current neutrino capture rates that are faster than antineutrino rates, thanks to the neutron-proton mass difference. As a result, the resulting nucleosynthetic conditions in the supernova neutrino-driven wind will tend to be proton-rich. If a sufficient excess of free protons is available following seed formation, along with a sizeable $\bar\nu_e$ flux to convert a fraction of the free protons into neutrons for $(n,p)$ and $(n,\gamma)$ reactions, then a $\nu p$ process will result.

\subsubsection{$s/k\sim 50$: $\nu p$-process nucleosynthesis}
\label{sec:vpprocess}

In the proton-rich neutrino-driven winds, the $\nu p$ process starts with seed nuclei around \ni56 (or possibly \ge64), synthesized in nuclear quasi-static equilibrium (QSE) at high temperatures ($T\geq 3$ GK). As the temperature drops below $\sim$3 GK, QSE freezes out and, in the absence of neutrinos, the nuclear reaction flow paths to higher $A$ via proton capture will be stalled by the long $\beta^{+}$-decay lifetimes of bottleneck (mainly even-even $N$=$Z$) nuclei. However, in the presence of neutrinos, antineutrino capture on free protons will produce a small increase in free neutrons. The bottlenecks can then be bypassed by neutron capture, allowing synthesis of increasingly heavy nuclei beyond iron \citep{Frohlich2006c, Pruet2006, Thielemann2010}. The $\nu p$-process yields depend on the hydrodynamic state of neutrino-driven winds, nuclear reaction rates, and the neutrino physics of the astrophysical site \citep{Wanajo2011, Arcones2012, Fujibayashi2015, Bliss2018b, Nishimura2019, Rauscher2019, Jin2020, Vincenzo2021, Sasaki2022, Sasaki:2023ysp}.
The symmetric neutrino parameters we adopt from Table~\ref{tab:neutrinomodels} result in initially proton-rich conditions with equilibrium electron fraction $Y_{e,\mathrm{eq}}= 0.634$. 

\begin{figure}[!htb]
	\centering
	\vspace{-5mm}
\includegraphics[width=12cm]{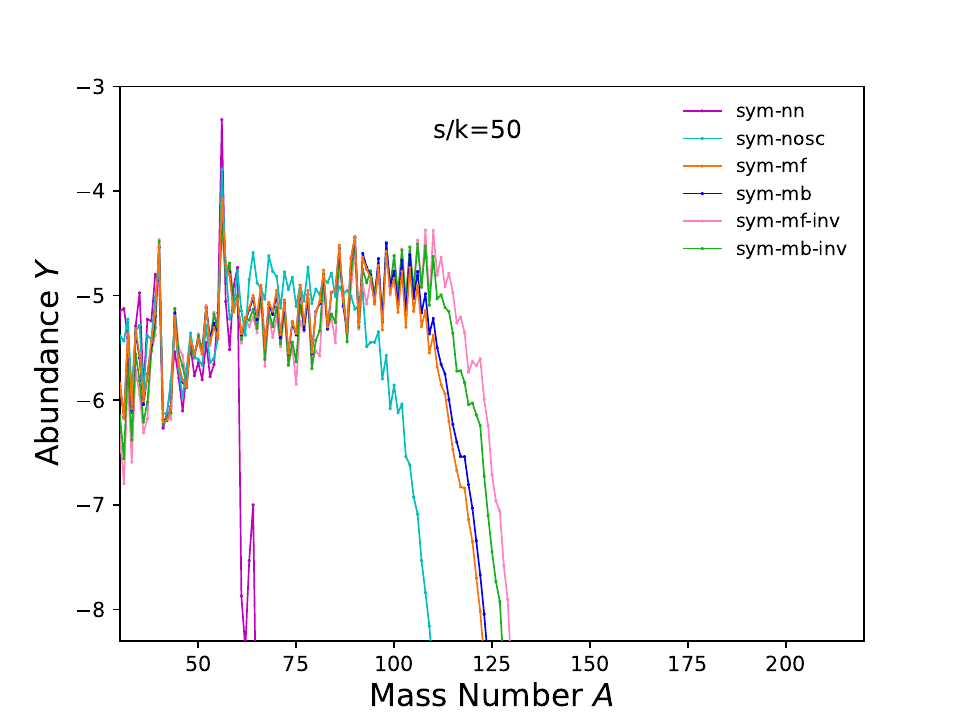}
\includegraphics[width=12cm]{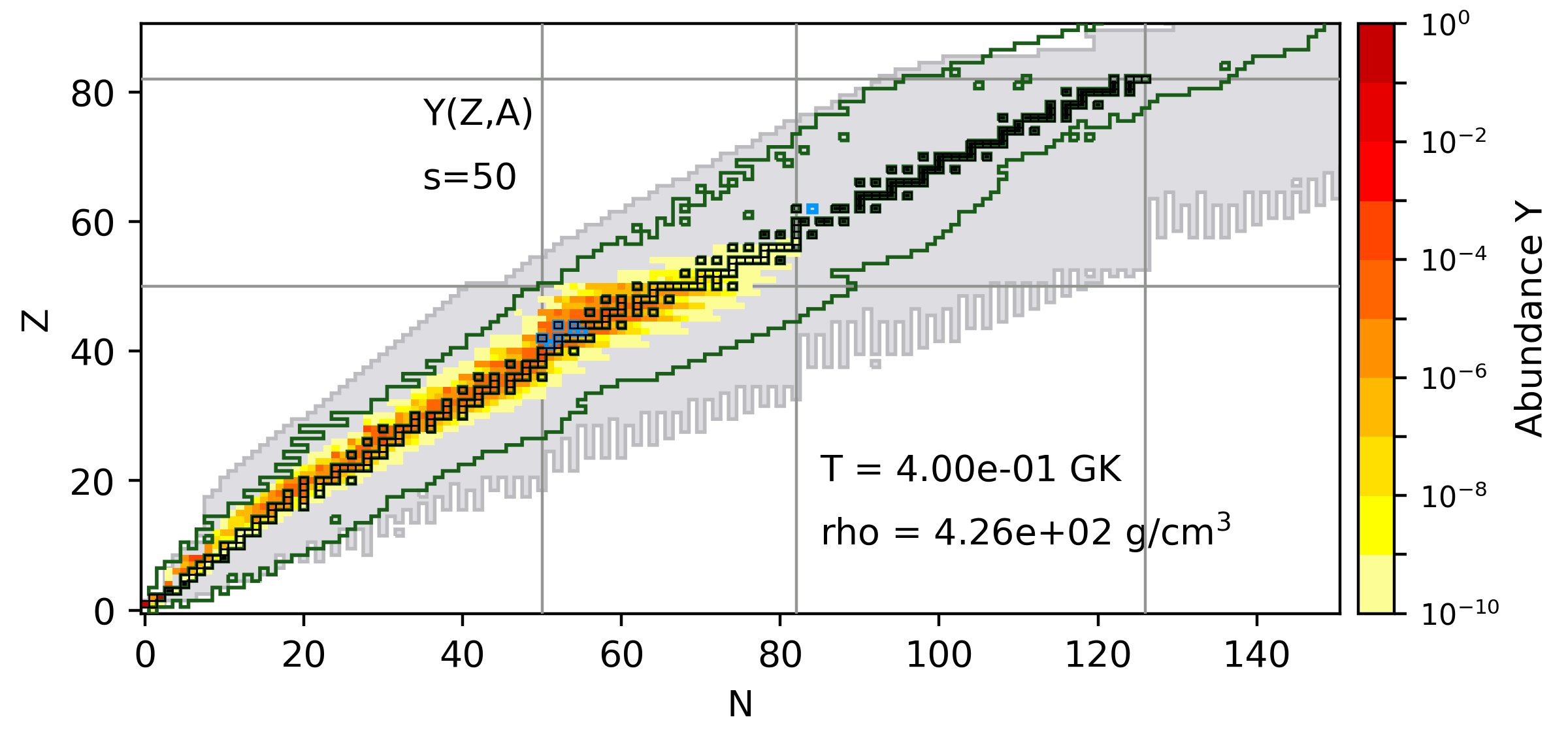}
	\caption{\textit{Top panel}: The final abundance patterns of simulations using the Wanajo2011 matter trajectory {with a fixed entropy-per-baryon $s/k = 50$ and} with symmetric \lq sym\rq\ neutrino energies {from Table~\ref{tab:neutrinomodels}} (purple: \lq nn\rq\ for no neutrinos involved for T $<$ 10 GK; cyan: \lq nosc\rq\ for no neutrino oscillations; orange: \lq mf\rq\ for mean-field neutrino oscillations; blue: \lq mb\rq\ for many-body neutrino oscillations), plotted as functions of the atomic mass number $A$. {Additionally, the abundances obtained using the symmetric neutrino distribution but with the inverted mass ordering are shown
as pink (mf) and green lines (mb).} \textit{Bottom panel}: Abundances in the $N$-$Z$ plane at the time when the nucleosynthesis pathway reaches its maximum extent in $A$ for the Wanajo2011 sym-mb case {in the normal mass ordering}. {The grey shaded region shows the full extent of the nuclear data used in the PRISM calculations, while the green line indicates the region of the nuclear chart for which some experimental nuclear data is available.}
	}
	\label{fig:wanajo10_sym}
\end{figure}

We follow the nucleosynthesis for $T<10$ GK in four distinct neutrino treatments, as elaborated in Sec.~\ref{sec:model}. 
To begin, we first test the effects of neutrino interactions for the Wanajo2011 model with $s/k=50$, which is a typical $\nu p$-process trajectory. The abundance pattern results are shown in Fig.~\ref{fig:wanajo10_sym}.
We can see that the final abundance pattern in the no-oscillations case shows typical $\nu p$-process features, with element synthesis up to $A\sim 100$, while the case with no neutrino interactions results in only iron-peak nuclei. Both cases are consistent with the results of \citet{Wanajo2011}. For the abundance yields in the cases with neutrino oscillations included, the enhanced neutrino capture rates shown in Fig.~\ref{fig:neutrino_duan_wanajo} create more neutrons throughout the $\nu p$ process. Thus the $\nu p$ process can proceed to heavier nuclei (Fig.~\ref{fig:wanajo10_sym}). As the material cools below temperatures required for proton capture, some of the remaining free protons continue to be converted to neutrons, producing a small amount of excess neutrons whose captures shift some material slightly \emph{neutron}-rich. The maximum extent of the reaction flow for the sym-mb case is illustrated in the bottom panel of Fig.~\ref{fig:wanajo10_sym}. Nuclei up to the $Z=50$ shell closure are populated, with a handful of species on the neutron-rich side of stability. The final abundances for the sym-mf and sym-mb cases remain those of a robust $\nu p$ process.

{We also tested the effect of neutrino interactions adopting the inverted mass ordering on the nucleosynthesis yields. In the two-flavor framework, which our calculations are based on, changing the mass ordering amounts to changing $\omega_p \rightarrow -\omega_p$, where $\omega_p = \Delta m^2/(2p)$ for neutrinos, and $-\Delta m^2/(2p)$ for antineutrinos. This difference leads to some differences in the oscillation behaviors. In the normal mass ordering, the extent of conversion is greater in the many-body case than in the mean-field case. In the inverted-ordering, the opposite is true: the mean-field case exhibits a greater degree of flavor-swapping than the many-body case. {Additionally, the inverted-ordering scenario results a greater degree of flavor conversion as a result of bipolar oscillations (e.g.,~\cite{Samuel:1995ri,Pastor:2001iu,Hannestad:2006nj,Duan:2006an}), resulting in higher neutrino average energies, higher capture rates, and consequently, a more robust synthesis of heavier elements. This behavior is illustrated in the top panel of Fig.~\ref{fig:wanajo10_sym}. The results obtained using the normal mass ordering, which are the main focus of this work, could thus be considered {conservative} in that sense.}}

\subsubsection{$s/k\gtrsim 100$: Emergence of a $\nu i$ process}
\label{sec:sym-vary}

The initial entropy of the astrophysical trajectory plays an important role in the resulting heavy element nucleosynthesis. Of most relevance here, entropy augments the ratio of free nucleons to seed nuclei at $T \sim 3$\,GK when QSE ends. This effect can be understood as follows: a higher entropy implies a lower density at a given temperature. Therefore, the reaction rates involved in the assembly of seed nuclei, which are strongly density-dependent, are lower at higher entropies, resulting in reduced seed formation -- and thereby a higher free-nucleon-to-seed ratio. 

\begin{figure}[!htb]
	\centering
	\vspace{-5mm}
\includegraphics[width=7.5cm]{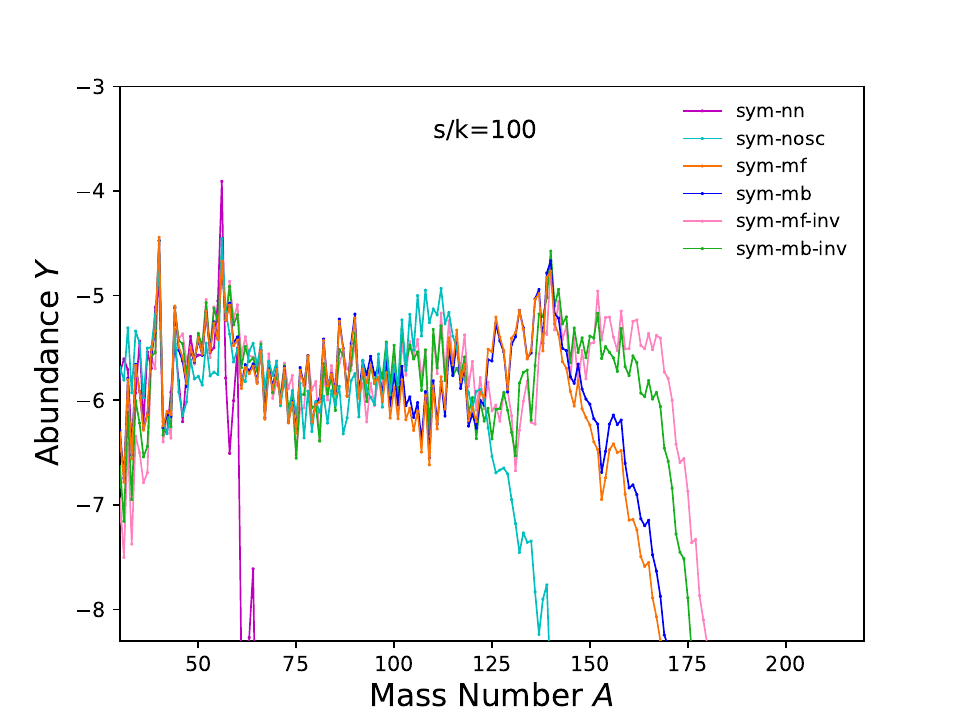}
\includegraphics[width=10cm]{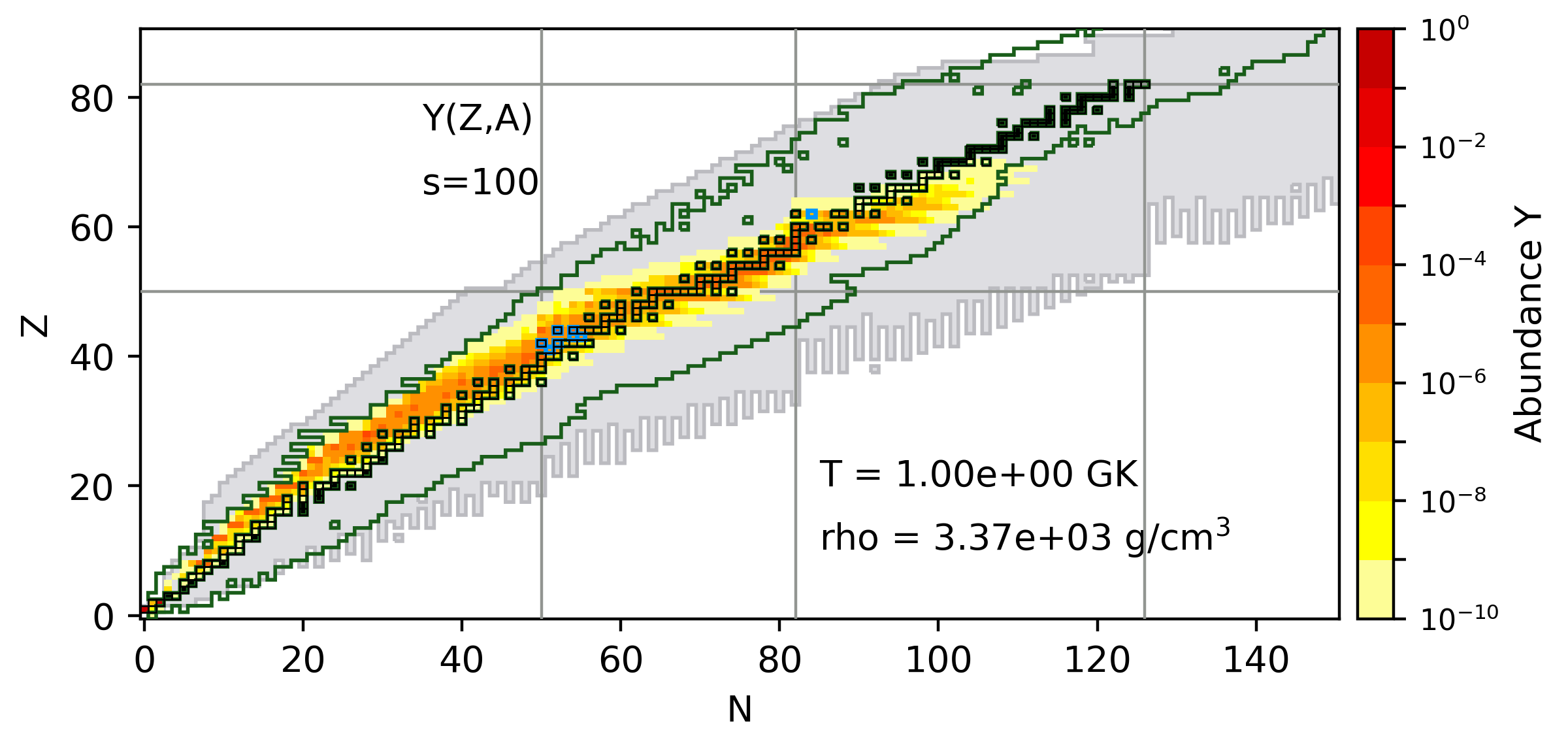}\\
\includegraphics[width=7.5cm]{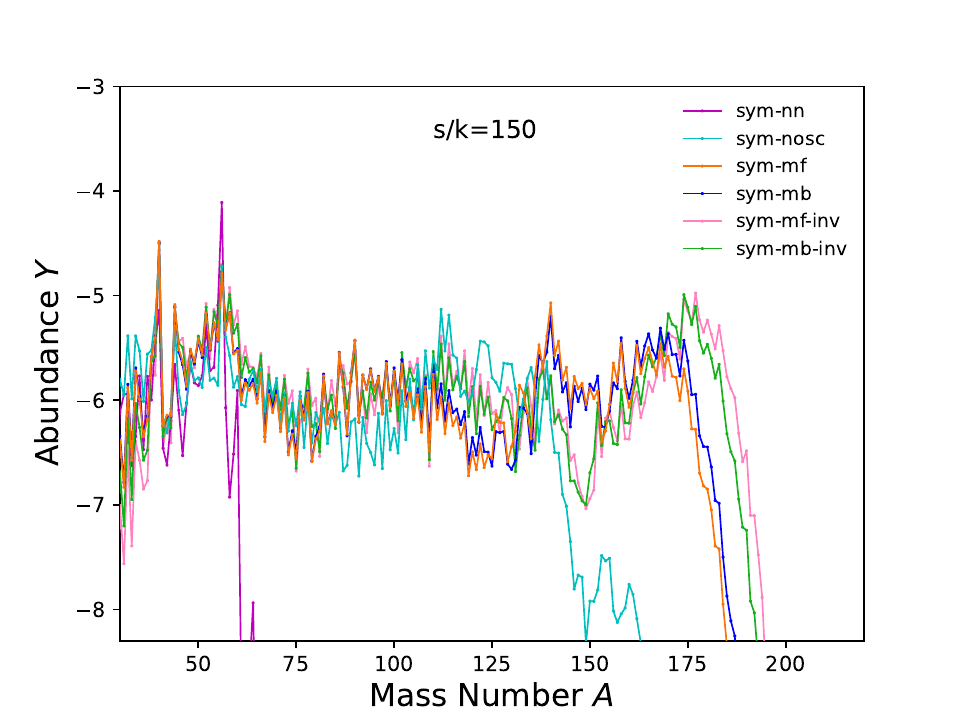}
\includegraphics[width=10cm]{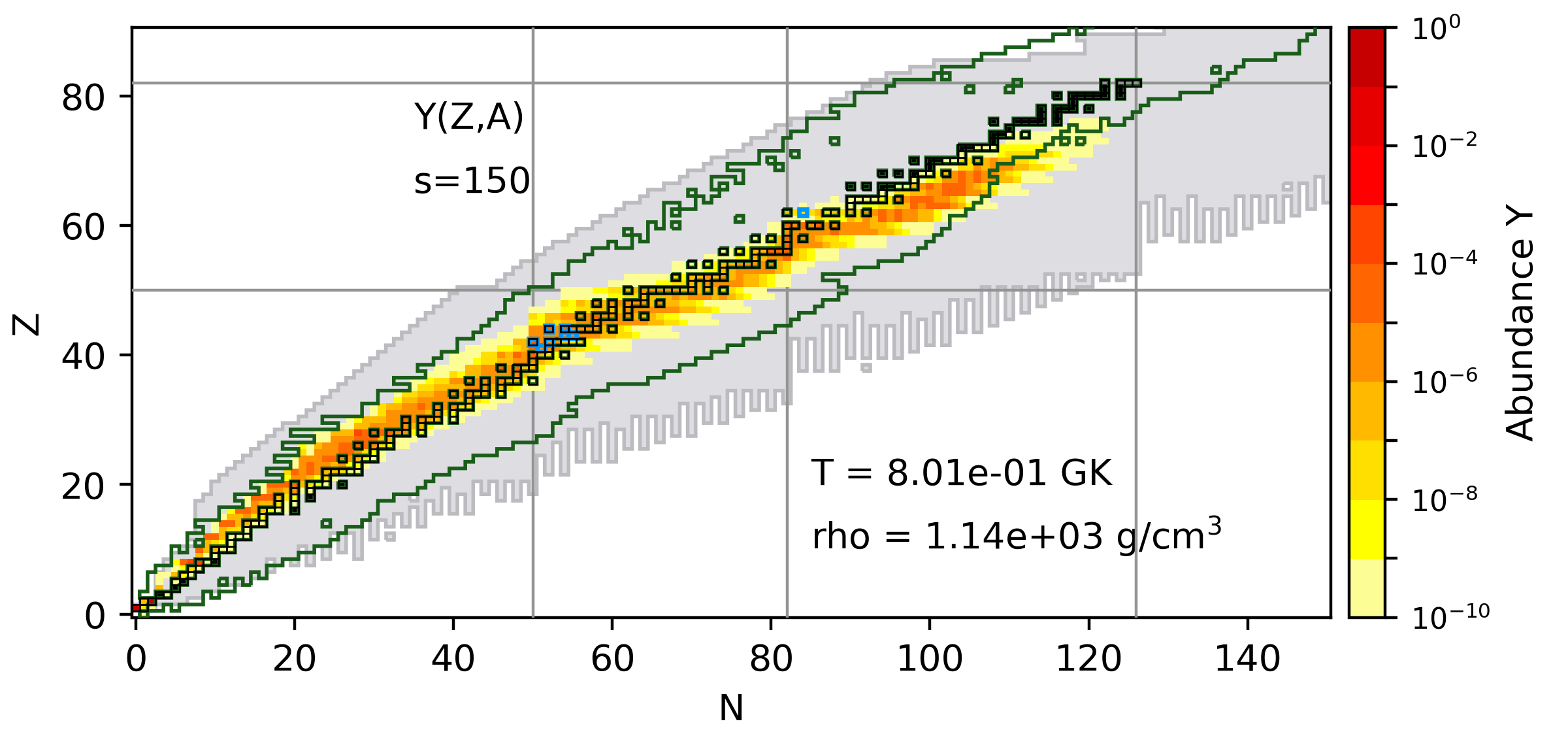}\\
	\caption{{\textit{Left panels}: The abundance patterns of the simulations using the Wanajo2011 matter trajectories {with fixed entropies-per-baryon $s/k = 100$ (\textit{top}) and $150$ (\textit{bottom}), and} with symmetric \lq sym\rq\ neutrino energies {from Table~\ref{tab:neutrinomodels}}, 
 plotted as functions of the atomic mass number $A$. {The line colors for the results 
 are the same is in Fig.~\ref{fig:wanajo10_sym}.}
 \textit{Right panels}: Abundances in the $N$-$Z$ plane at the time when the nucleosynthesis pathway reaches its maximum extent in $A${, using the many-body neutrino calculation in the normal mass ordering}.
	}}
	\label{fig:wanajo10-sym-vary}
\end{figure}

We examined variations in the initial entropy of the Wanajo2011 trajectories from $s/k=50$ to 100 and 150, with the same set of four distinct neutrino treatments. The results for $s/k = 100$ and 150 are shown in Fig.~\ref{fig:wanajo10-sym-vary}. 
When the entropy is increased to $s/k=100$, all cases that include neutrino interactions generate heavier elements compared to the $s/k=50$ example discussed above. Furthermore, the higher free proton-to-seed ratio provides greater leverage for the influence of neutrino oscillations. The higher neutrino capture rates of the sym-mf and sym-mb cases produce dramatically different nucleosynthetic outcomes from those of the sym-nosc case, with the sym-mb case producing the heaviest nuclear species. At $s/k=150$, this influence is even more significant. Interestingly, in these examples, there are even more free protons available at low temperatures. Consequently, more neutrons are available at late times, such that neutron capture becomes not only a means to bypass certain slow $\beta^+$ half-lives but actually robust enough to drive the reaction flow across the valley of stability and over to the neutron-rich side for $Z \gtrsim 50$. The nucleosynthesis path in the right panels of Fig.~\ref{fig:wanajo10-sym-vary} shows that, for sym-mb cases, the reaction flow shifts to the neutron-rich side of the valley of stability at high $A$ and synthesis proceeds to heavy nuclei beyond the $Z=50$ and $N=82$ shell closures. In the sym-mb case, the final abundances reach $A\sim 160$ for $s/k=100$ and $A\sim 190$ for $s/k=150$.

\begin{figure}[!htb]
	\centering
	\vspace{-5mm}
\includegraphics[width=8.5cm]{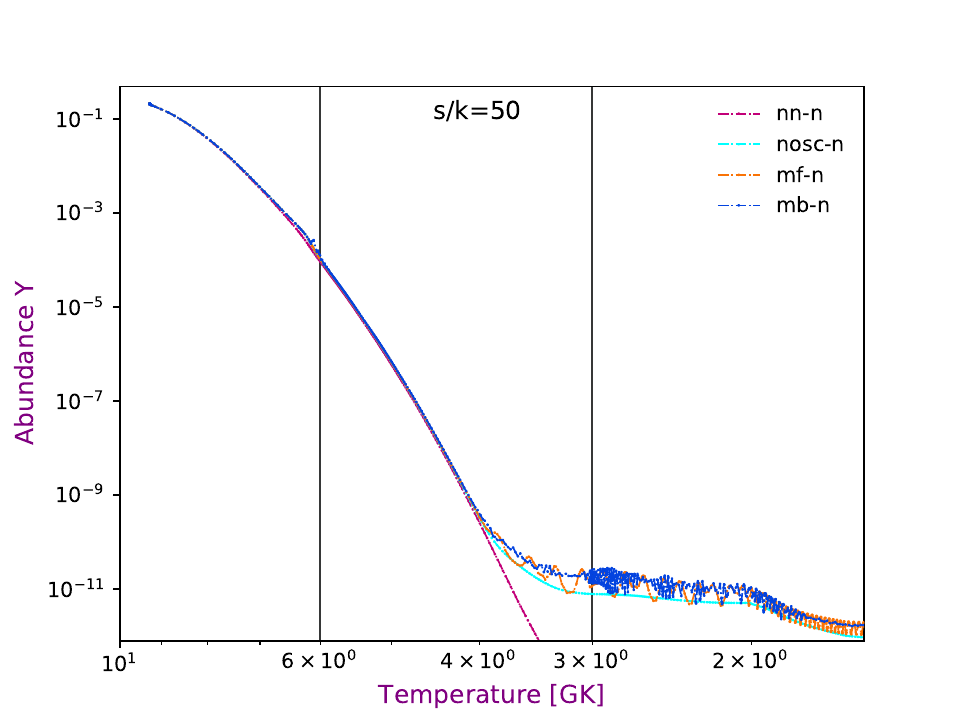}
\includegraphics[width=8.5cm]{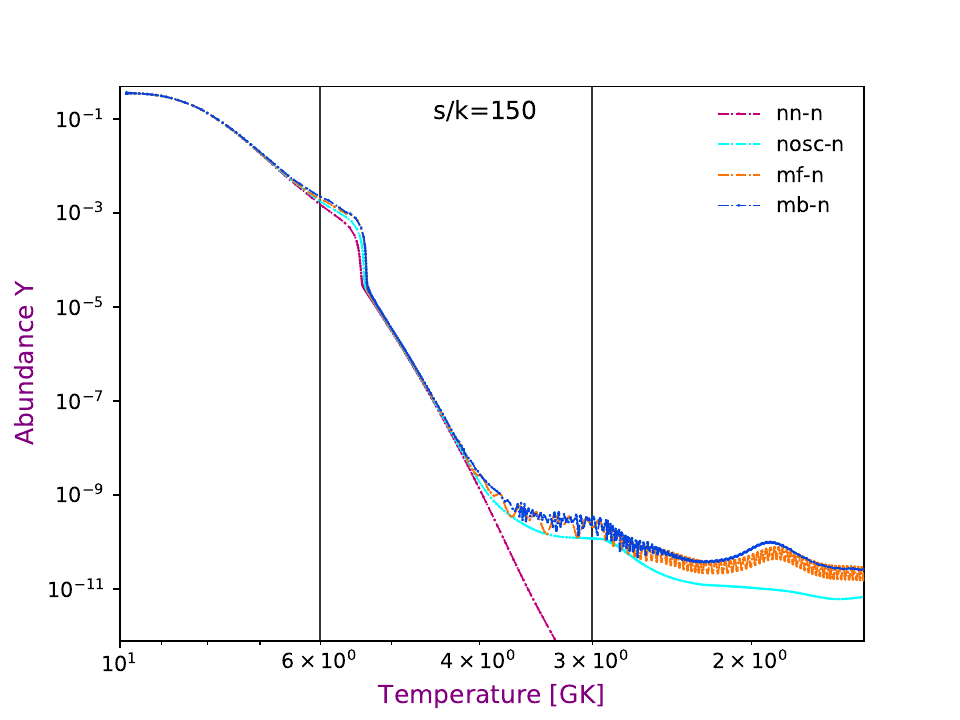}\\
\includegraphics[width=8.5cm]{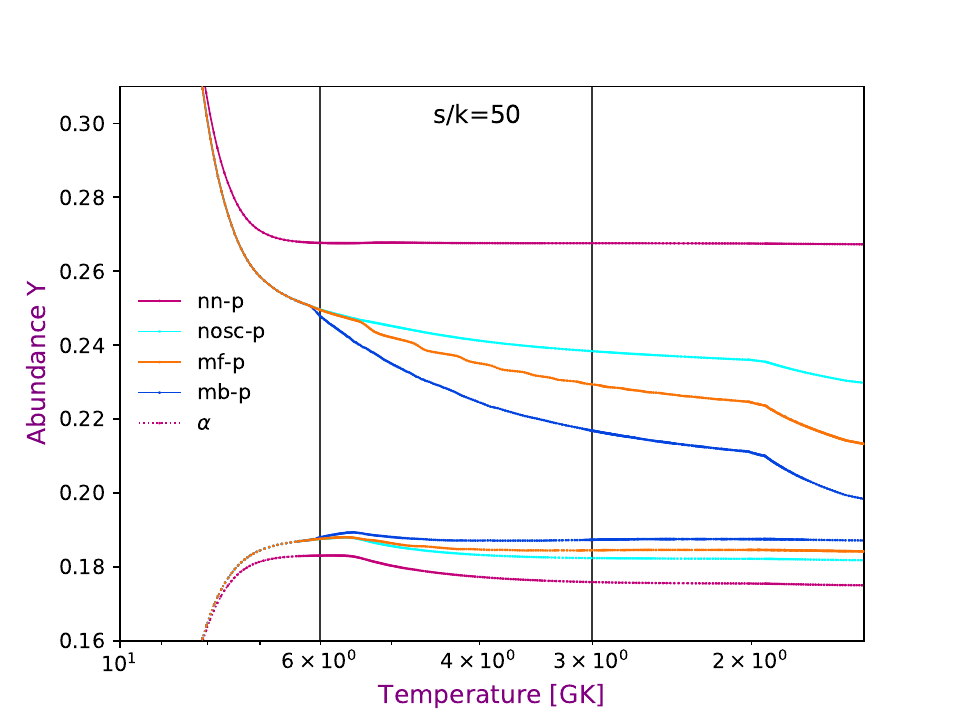}
\includegraphics[width=8.5cm]{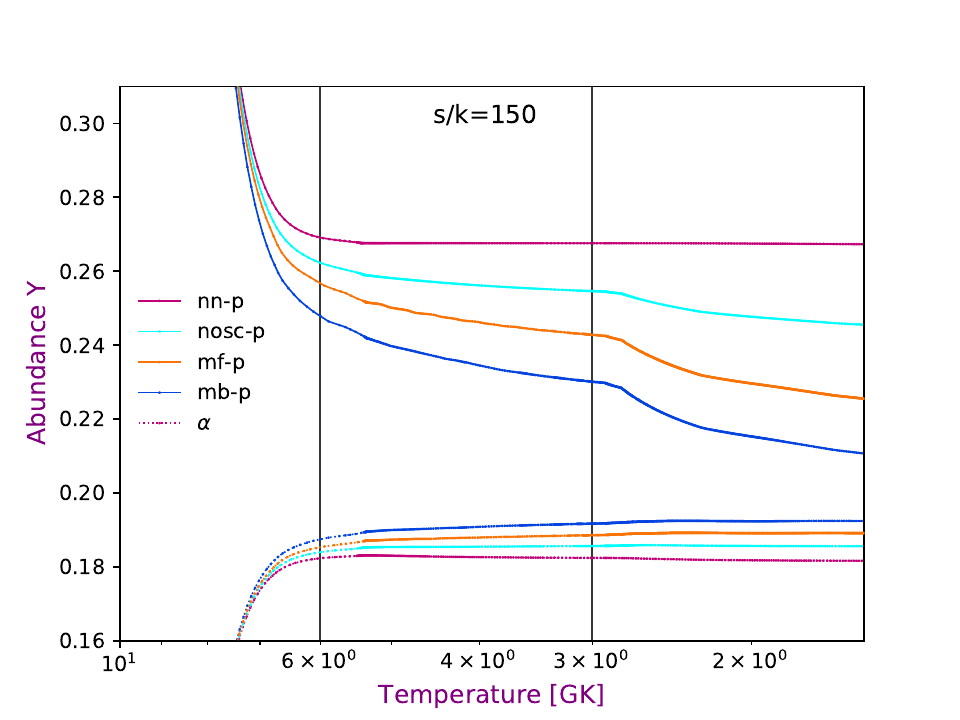}\\
	\caption{Neutron (top) and proton and alpha particle (bottom) abundances versus temperature for the Wanajo2011 $s/k=50$ (left) and $s/k=150$ (right) trajectories and symmetric neutrino calculations (line colors as in Fig.~\ref{fig:wanajo10_sym}).}
	\label{fig:wanajo2010_sym_evol}
\end{figure}

To understand the mechanism by which this transition to neutron capture nucleosynthesis occurs, we show the temperature evolution of the abundances of neutrons (top panel, log scale) and protons (bottom panel, linear scale) in Fig.~\ref{fig:wanajo2010_sym_evol}. Without neutrino interactions on free nucleons, iron group nuclei are synthesized and additional proton capture is blocked by the slow $\beta$ bottlenecks, shown in the red lines of each of the panels of Fig.~\ref{fig:wanajo2010_sym_evol}. 
When neutrino interactions are included (without oscillations), shown by the cyan lines of Fig.~\ref{fig:wanajo2010_sym_evol}, the proton abundance continues to decrease as the temperature drops, due to conversion of protons to neutrons by neutrinos and the resulting increase in proton captures to higher $A$. Both of these effects are enhanced by neutrino oscillations, with the sym-mb case (blue lines) showing a larger effect than sym-mf (orange lines). Comparing the $s/k=150$ examples to $s/k=50$, we see the influence of the oscillations at an earlier stage of the nucleosynthesis, since we take the onset of oscillations to occur at a fixed radius, which corresponds to a higher temperature for the $s/k=150$ outflow. The neutrons produced at $T>6$ GK end up bound into alpha particles, with enhanced production of alpha particles for the sym-mb and sym-mf cases. Still, the higher entropy ensures that fewer seed nuclei are assembled from the alpha particles, so the free proton-to-seed ratio remains high. The production of neutrons and their subsequent capture is therefore enhanced. This effect allows late time, low ($T<1.5$ GK) temperature capture of neutrons and reaction flow to heavier, neutron-rich nuclei. {As with the $s/k=50$ case, the effect is stronger for the inverted mass ordering calculations than for normal ordering.}

The neutron-rich portions of the nucleosynthetic pathways illustrated in Fig.~\ref{fig:wanajo10-sym-vary} are similar to those that characterize another nucleosynthesis process: the $i$ process. The $i$ process is an intermediate neutron capture process that results from neutron-rich conditions in which the neutron number density is greater than that of the slow neutron capture ($s$) process but significantly less than required for an $r$ process, e.g.\ $N_{n}\sim 10^{13}-10^{15}$ cm$^{-3}$ \citep{CowanRose1977}. It has been suggested that neutron number densities within this range can be attained in, for example, the helium flash of asymptotic giant branch (AGB) and post-AGB stars \citep{CowanRose1977,Malaney1986,Jorissen1989,Cristallo+2016} and rapidly-accreting white dwarfs \citep{denissenkov+2017,Denissenkov+2019}, and there is tentative evidence that points to $i$-process contributions to some unusual stellar abundance patterns \citep{Dardelet2014,Hampel+2019}.  The process that produces this $i$-process-like nucleosynthetic pathway here is clearly distinct from an $i$ process that occurs in mildly neutron-rich conditions. Instead, here the heavy elements are created via neutron capture where the neutrons are a product of neutrino interactions on excess protons \textit{in a proton-rich wind}. The suggestion that a robust $\nu p$ process could shift to neutron-rich species at late times was first pointed out in \cite{Wanajo2011}, \cite{Arcones2012}, and also \cite{Nishimura2019}, though the details were not explored. A similar process (with smaller maximum mass number $\sim160$) was also observed in a hypernova neutrino-driven wind with higher neutrino luminosities $L_{\nu}\sim 10^{53}$ erg/s \citep{Fujibayashi2015}. We find neutrino oscillations amplify the shift from proton-rich to neutron-rich nucleosynthesis and can result in a full intermediate neutron capture process. We dub this novel nucleosynthesis process a ``$\nu i$ process''.

\subsubsection{$s/k\sim 200$: $\nu i$-process nucleosynthesis}
\label{sec:viprocess}

Section~\ref{sec:sym-vary} shows that the influence of different neutrino treatments on proton-rich nucleosynthesis grows with the entropy of the matter trajectory. Here we adopt a supernova neutrino-driven wind trajectory with faster evolution and higher entropy ($s/k=200$) from Duan2011 to test the potential extent of the $\nu i$ process in a more extreme case.

\begin{figure}[!htb]
	\centering
	\vspace{-5mm}
\includegraphics[width=12cm]{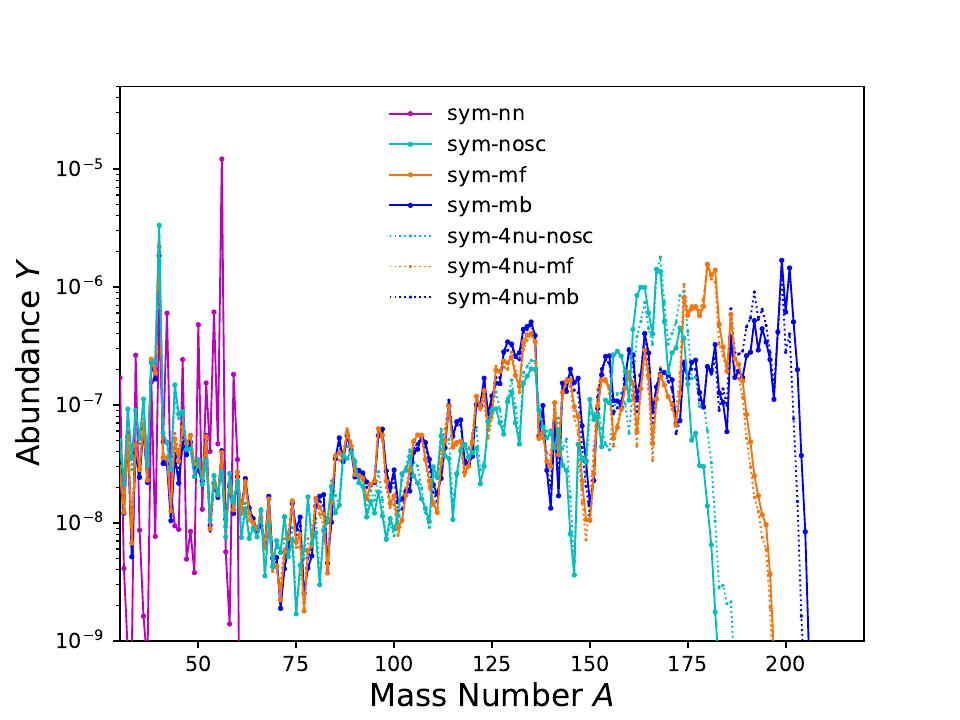}\\
	\caption{The abundance patterns of model Duan2011 with different symmetric neutrino calculations (purple: nn for no neutrinos involved for T $<$ 10 GK; cyan: nosc for no neutrino oscillations; orange: mf for mean-field neutrino oscillations; blue: mb for many-body neutrino oscillations) for 2nu (2 $\nu$ + 2 $\bar\nu$; solid lines) and 4nu (4 $\nu$ + 4 $\bar\nu$; dotted lines), plotted as functions of the atomic weight $A$. 
	}
	\label{fig:duan2011_sym}
\end{figure}

\begin{figure}[!htb]
	\centering
\includegraphics[width=8.5cm]{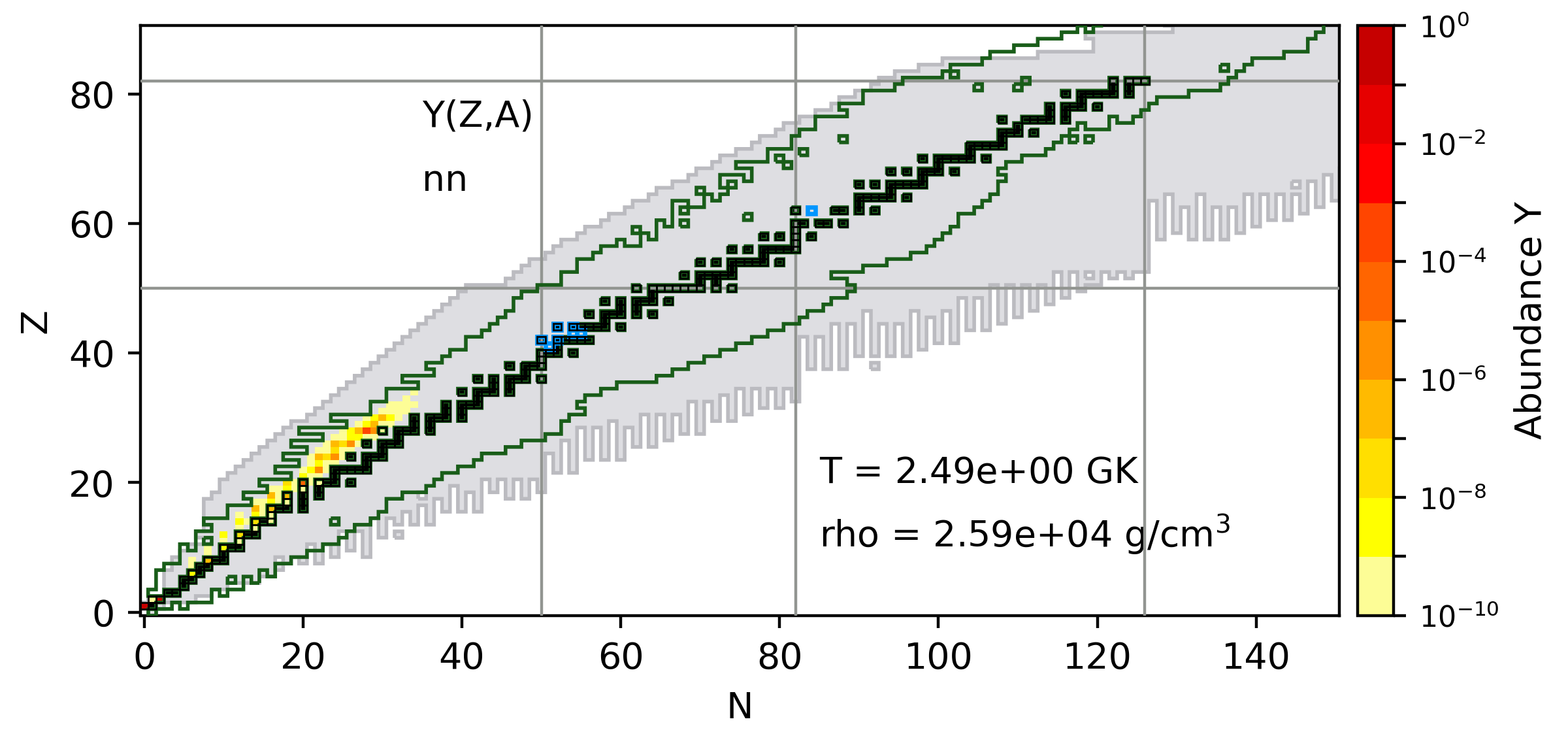}
\includegraphics[width=8.5cm]{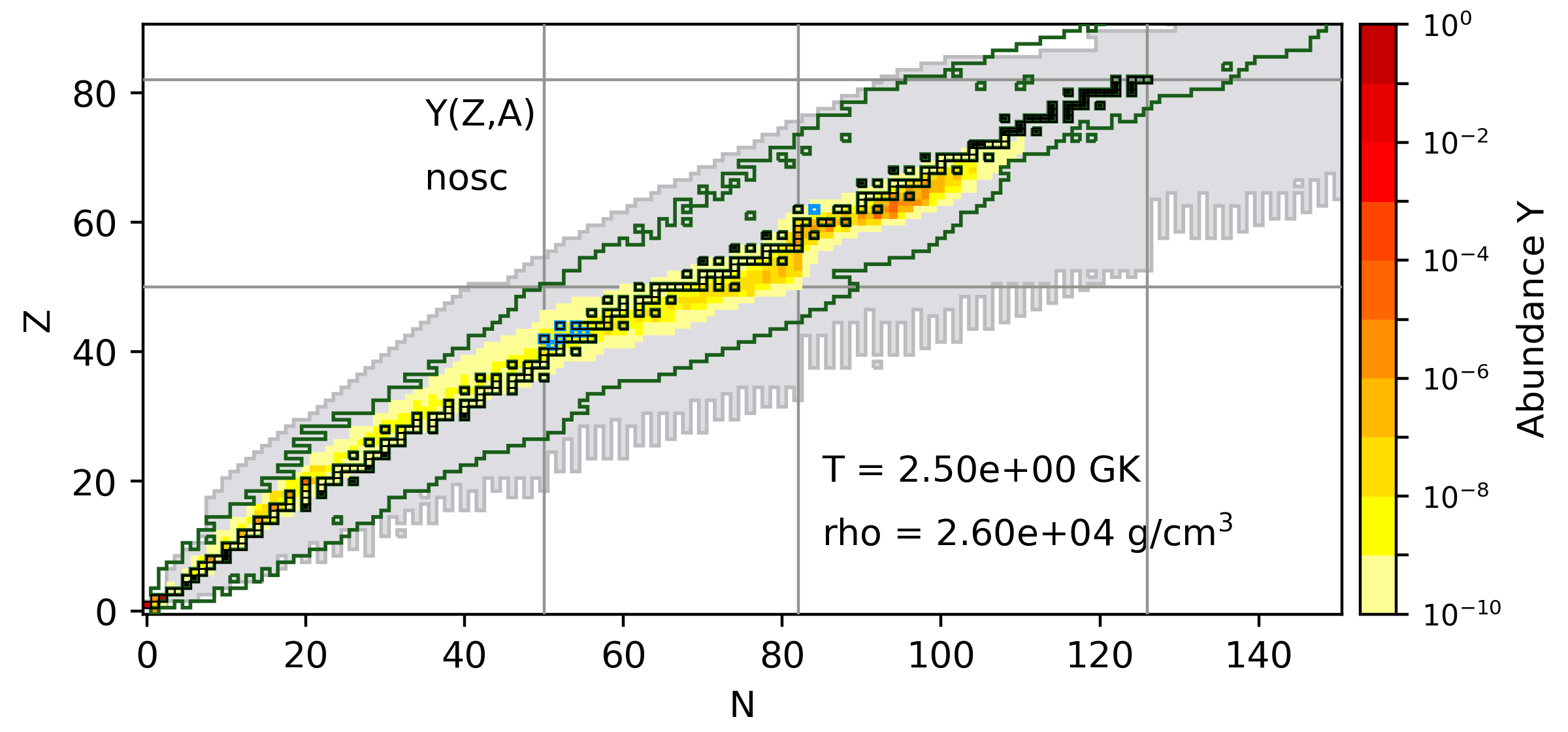}\\
\includegraphics[width=8.5cm]{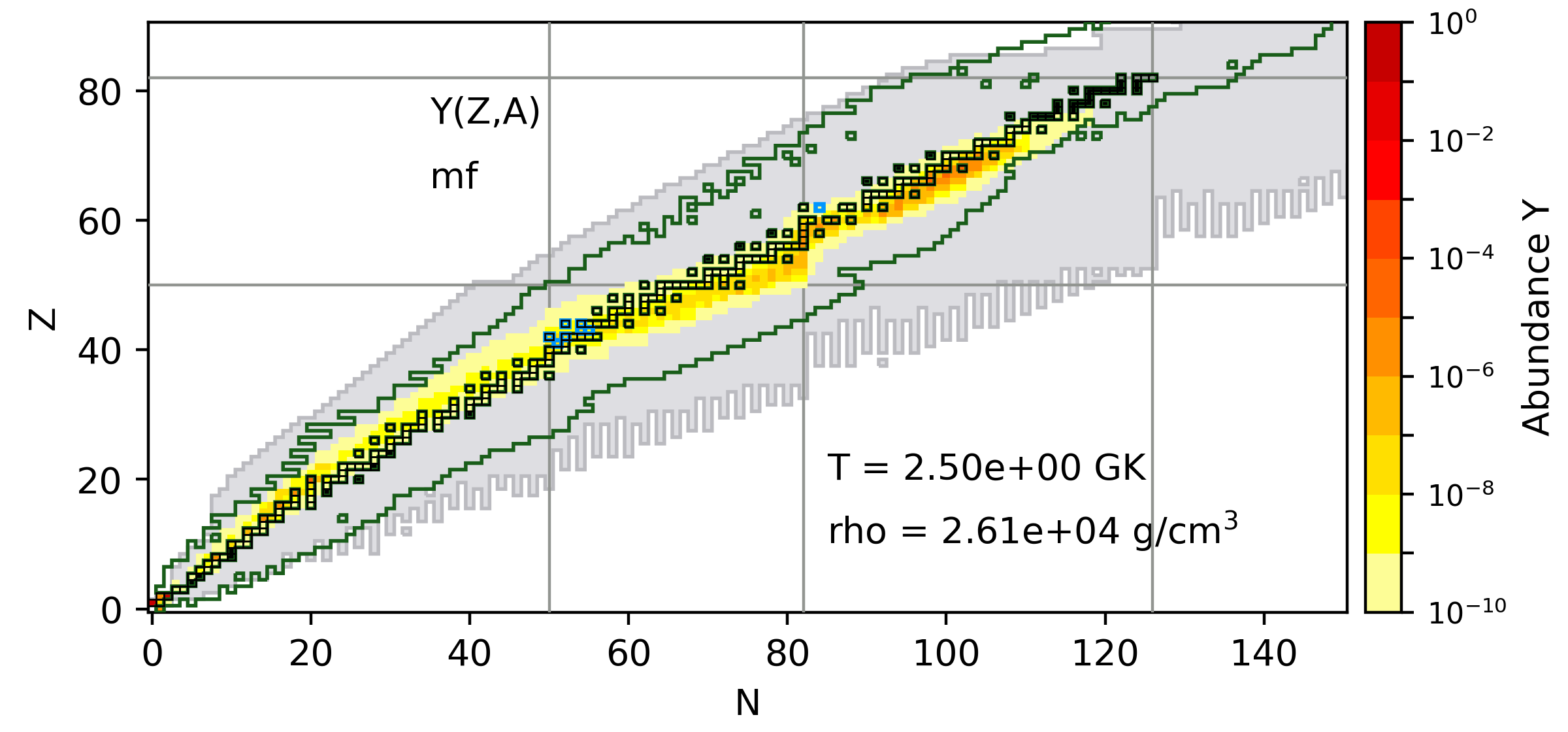}
\includegraphics[width=8.5cm]{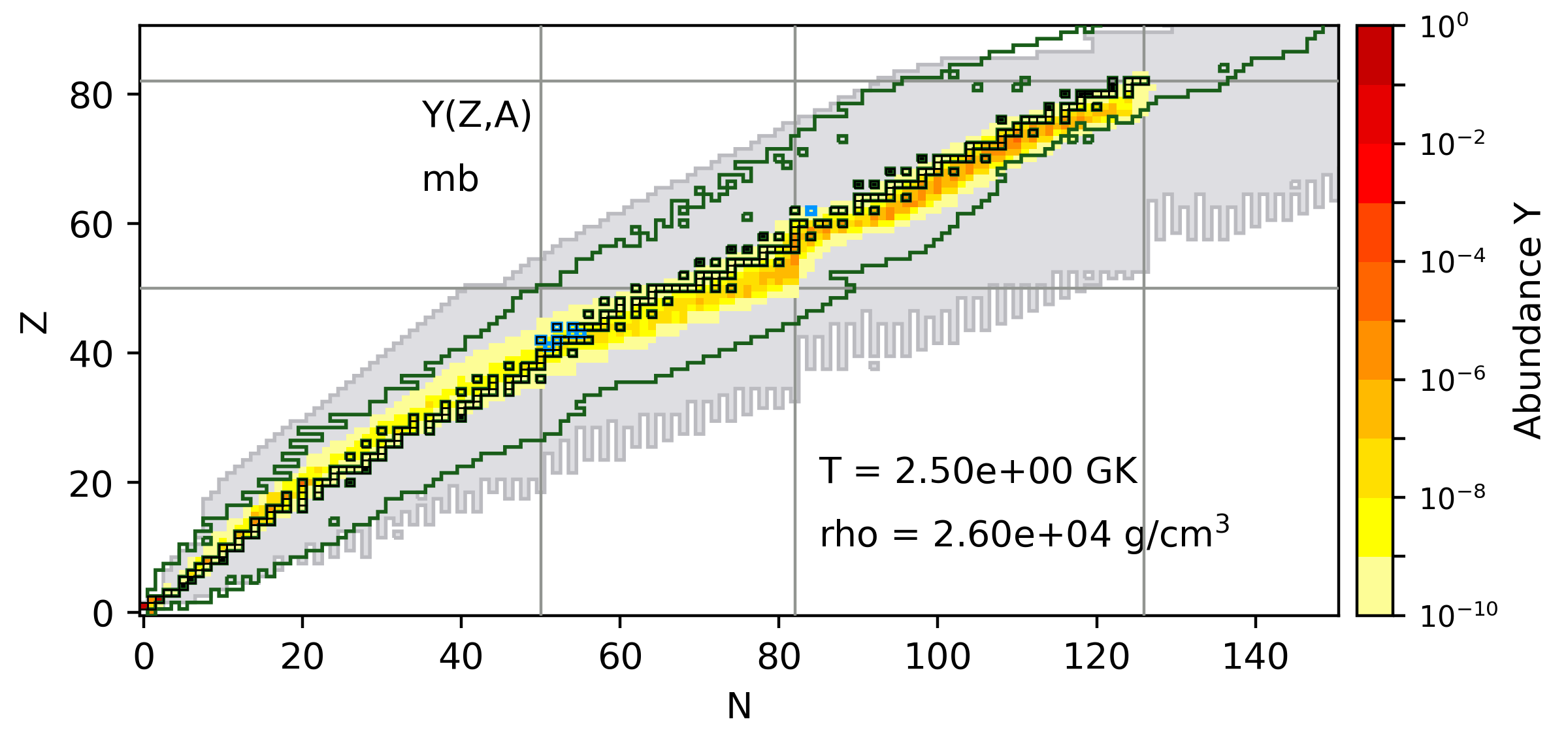}\\
	\caption{The nucleosynthesis paths of the Duan2011 matter trajectory at $T=2.5$ GK: sym-nn (upper left), sym-nosc (upper right), sym-mf (lower left), and sym-mb (lower right). 	}
	\label{fig:duan2011_sym_path}
\end{figure}

The abundance pattern results for the Duan2011 trajectory with the four distinct neutrino treatments are shown in Fig.~\ref{fig:duan2011_sym}. We can see that, compared to the Wanajo2011 $s/k=150$ case, there is even more robust production of heavy species and a clearing out of light species via neutron capture, which proceeds to high $A$ ($\sim 205$ for mb case). Also, the differences in the nucleosynthetic outcomes among the neutrino treatments are enhanced, with the sym-mb showing the most robust production of heavy species. This behavior is further illustrated in Fig.~\ref{fig:duan2011_sym_path}, which shows the nucleosynthetic pathway at the time when the crossover to the neutron-rich side of stability occurs (at temperature $T\sim2.5$ GK). The sym-mb case shows the greatest extent in $A$, reaching almost up to the stability gap.

Given the dramatic difference between the sym-mf and sym-mb results in this scenario, we explore the possibility that the difference manifested in this extreme case is due to the uncertainties inherent in our simplified many-body neutrino treatment. We therefore also examined a many-body calculation with increased neutrino numbers (4nu: 4 $\nu$ + 4 $\bar\nu$). A comparison of the nucleosynthesis results with the default 2nu (2 $\nu$ + 2 $\bar\nu$) calculation are shown in Fig.~\ref{fig:duan2011_sym}. We can see that the abundance patterns with different neutrino treatments for the default 2nu calculations and the 4nu case are very similar. This trend suggests that the differences we see in the sym-mf and sym-mb results are due to the distinct oscillation patterns in the two cases in our setup, a difference which is amplified by the high entropy/fast evolution of the Duan2011 matter trajectory.

\begin{figure}[!htb]
	\centering
\includegraphics[width=8.5cm]{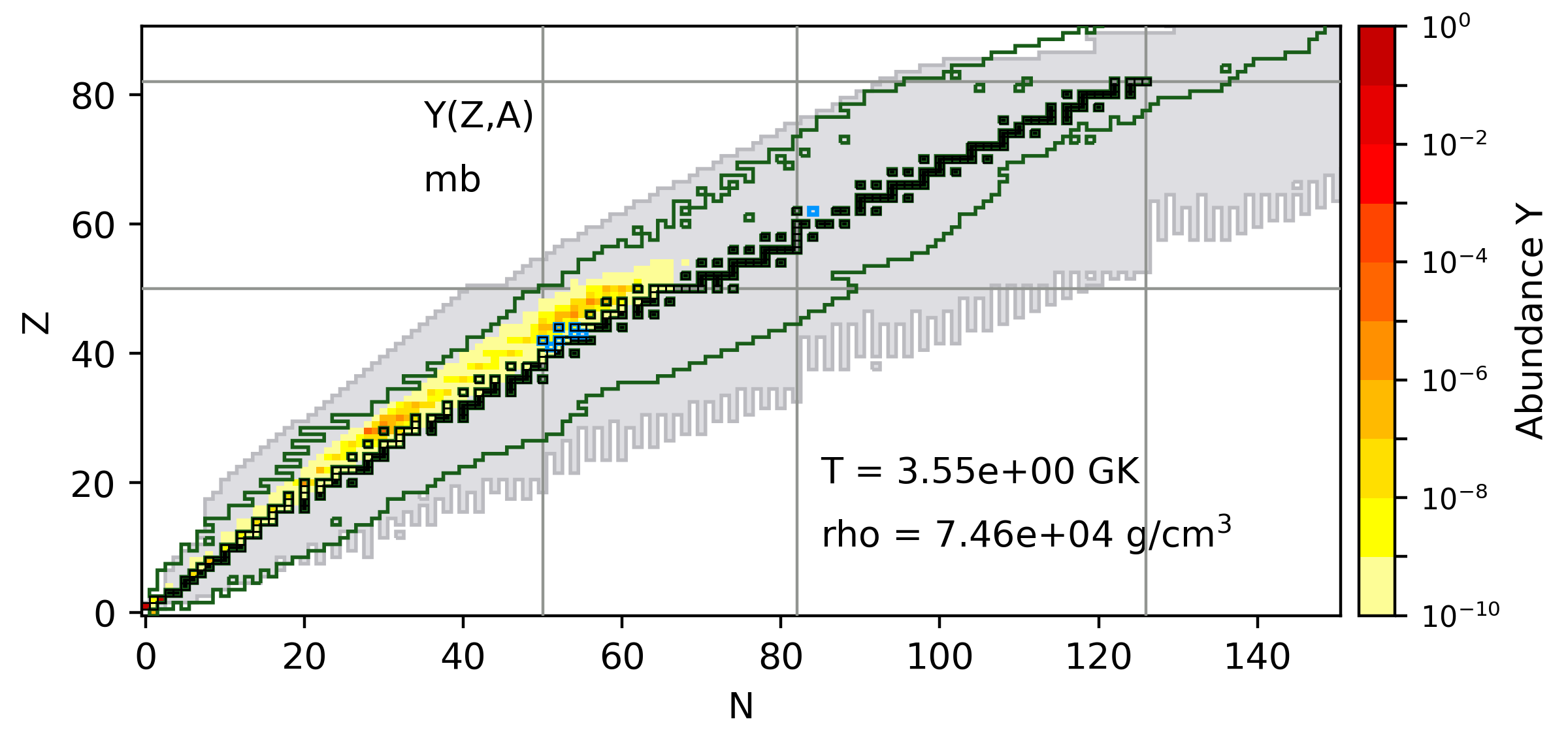}
\includegraphics[width=8.5cm]{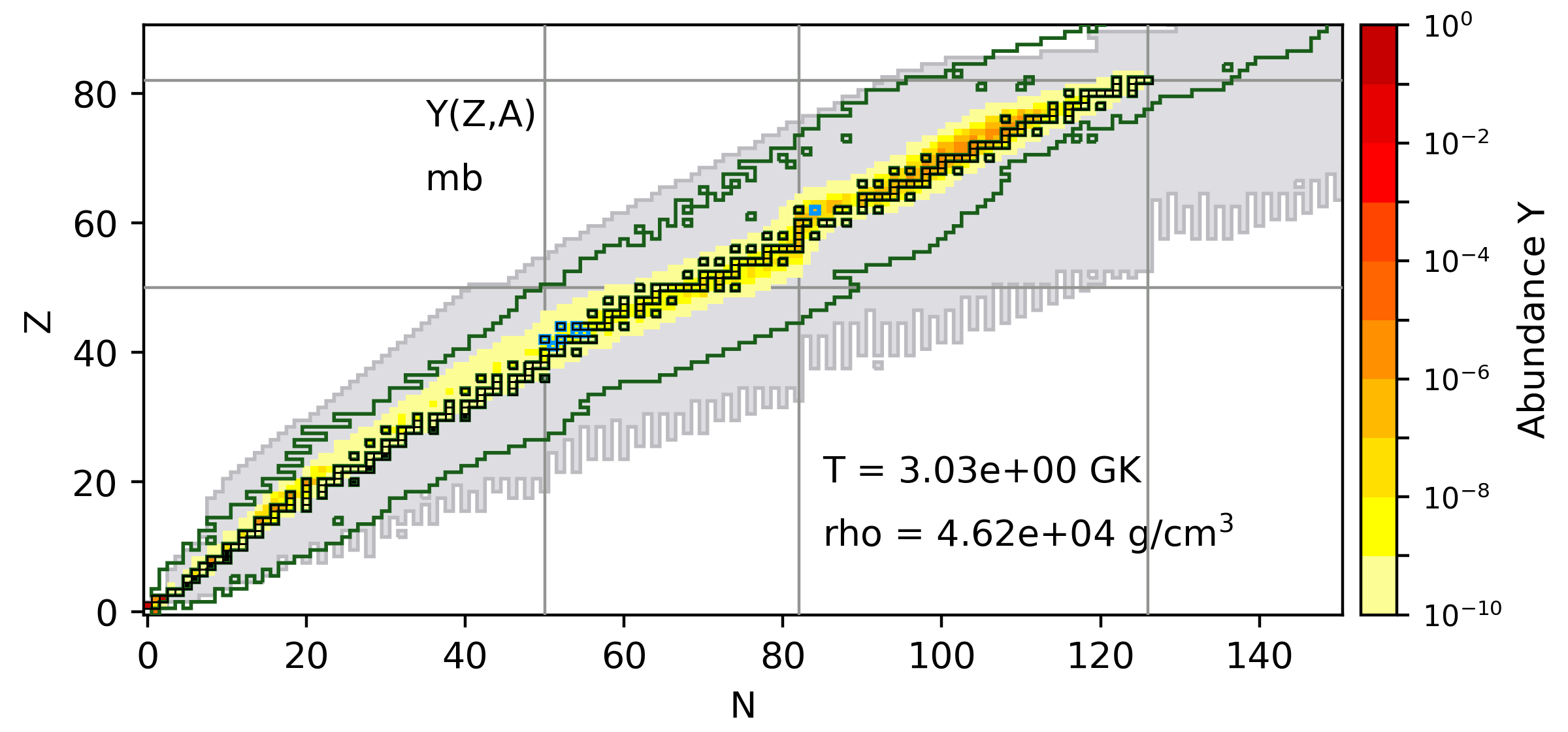}
\includegraphics[width=8.5cm]{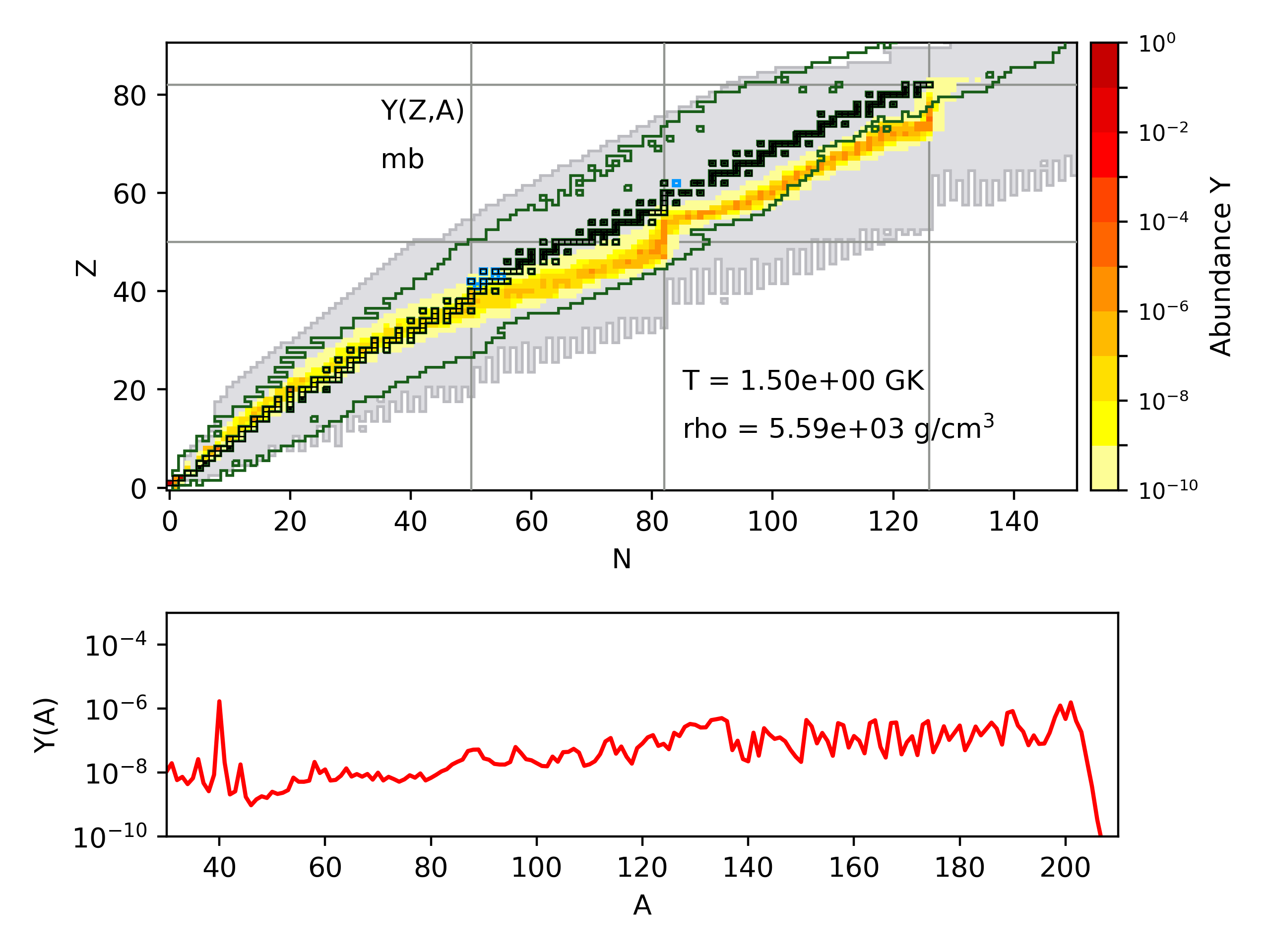}
\includegraphics[width=8.5cm]{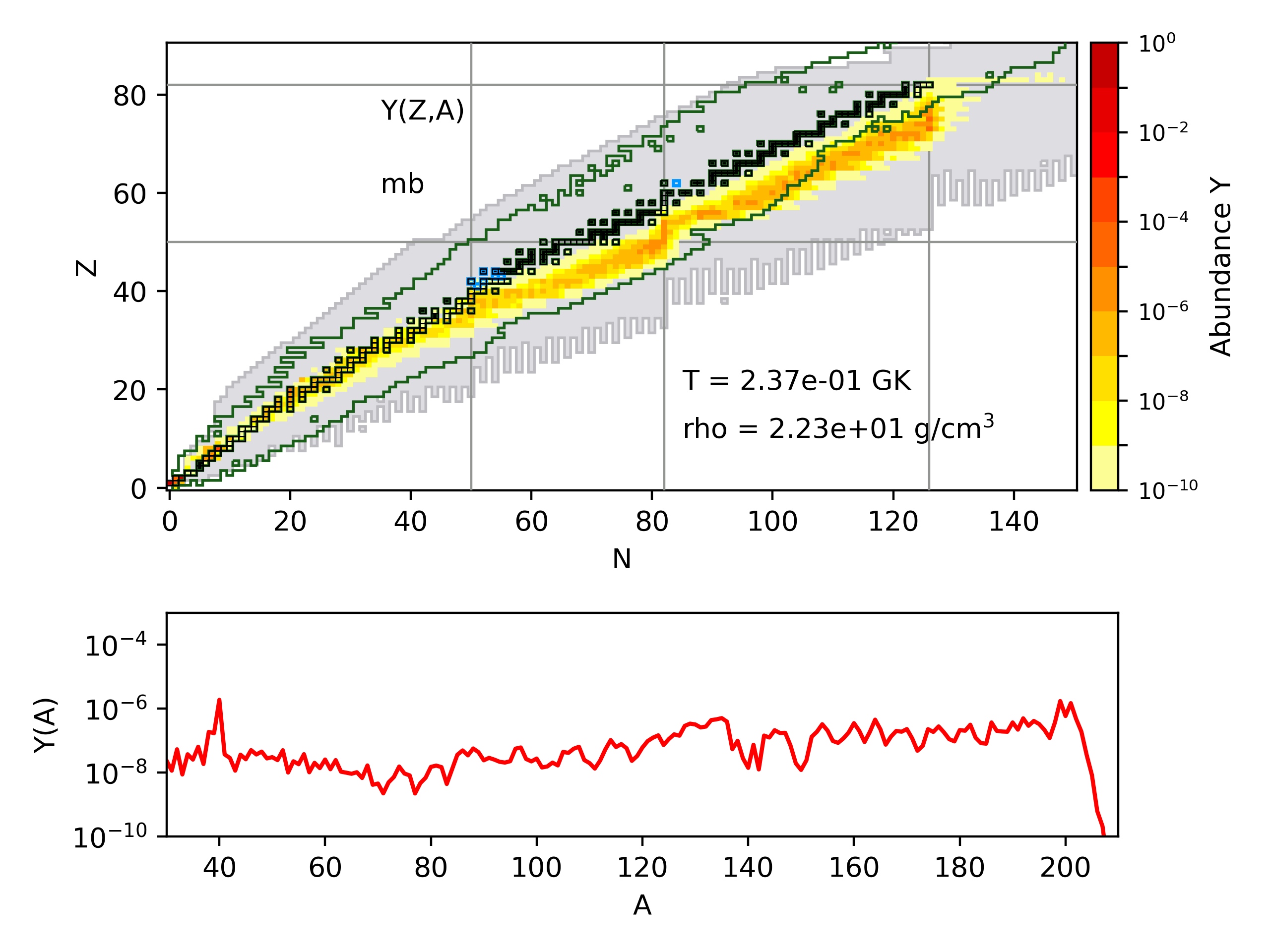}
	\caption{The nucleosynthesis paths of the Duan2011 matter trajectory with sym-mb neutrino calculations, shown at distinct phases of the nucleosynthesis: when entirely proton-rich (top left), as the pathway crosses the valley of stability (top right), at the furthest extent on the neutron-rich side (middle left), and at late time when most nuclei are neutron-rich (middle right). The bottom row shows the abundances as a function of $A$ for the latter two times. }
	\label{fig:duan2011_sym_mb}
\end{figure}

We explore the details of the sym-mb case in Fig.~\ref{fig:duan2011_sym_mb}, which shows the evolution of the nucleosynthetic pathway as a function of time. At early times/high temperatures ($T>3.5$ GK), the path is on the proton-rich side of stability, as for a typical $\nu p$ process. By $T\sim 3$ GK, the production of neutrons by neutrino interaction and their subsequent capture via ($n$,$\gamma$), ($n$,$p$), and ($n$,$\alpha$) have shifted the pathway to the valley of stability. At lower temperatures, $T<2$ GK, charged particle capture rates slow and neutron capture takes over in earnest. 

\begin{figure}[!htb]
	\centering
\includegraphics[width=15cm]{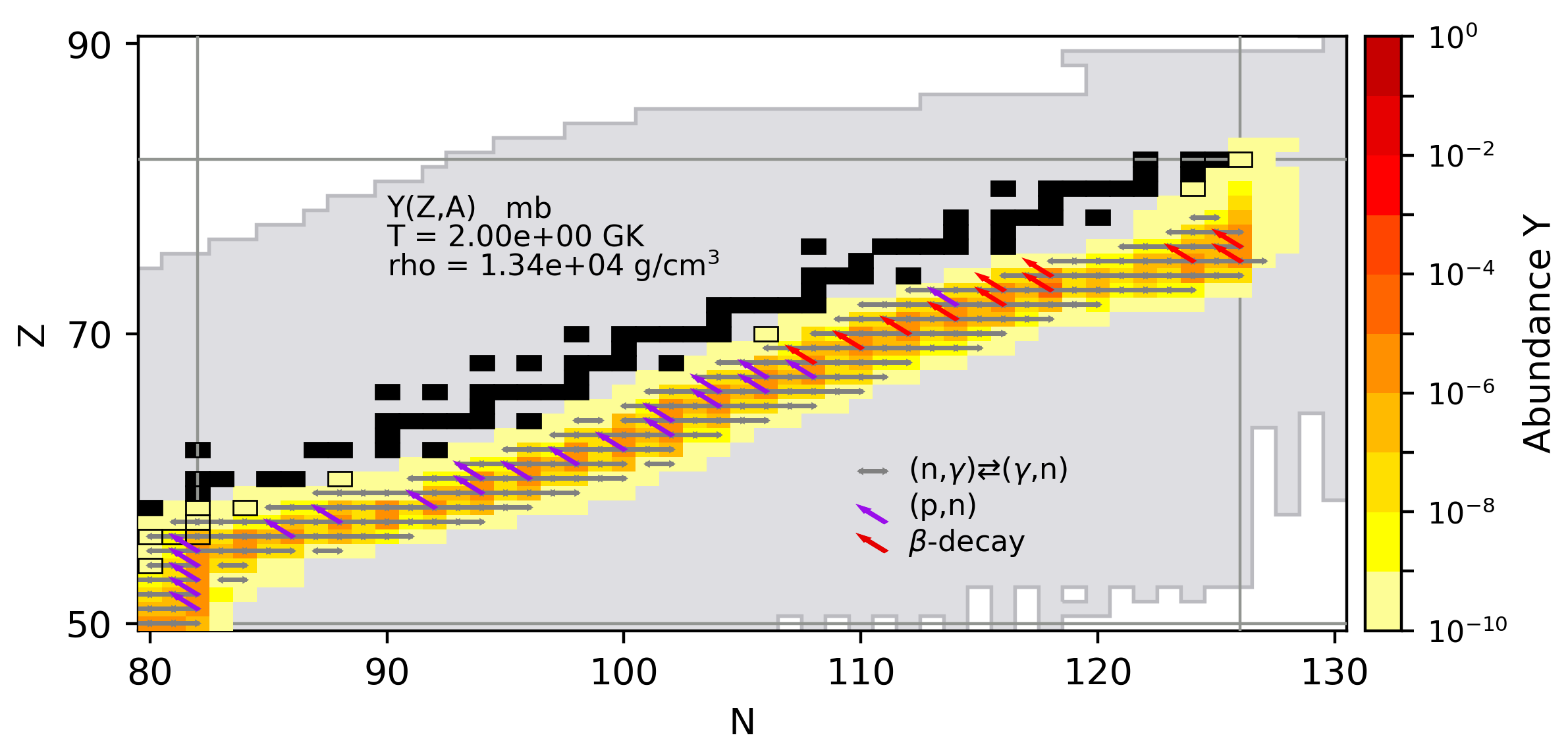}
	\caption{{The dominant nuclear reaction flows of the Duan2011 matter trajectory with sym-mb neutrino calculations, shown at $T=2$ GK. The primary reaction channels are neutron captures ($n$,$\gamma$) and  photodissociations ($\gamma$,$n$), which are roughly in equilibrium, denoted by the grey two-way arrows. The red arrows ($\beta^{-}$ decay) or purple arrows ($p$,$n$) show the dominant charge-changing reaction for the most abundant nuclear species along each isotopic chain. The red-yellow color bar denotes the abundances of each ($Z$,$N$) at $T=2$ GK.}}
	\label{fig:path_duan2011_sym_mb}
\end{figure}

{The dominant nuclear reaction flows at $T=2$\,GK are shown in Fig.~\ref{fig:path_duan2011_sym_mb}. Similar to a `traditional' $i$ process, depicted in, for example, \cite{Choplin2020,Choplin2021a,Choplin2021b,Choplin2022}, the predominant reaction is neutron capture. Here, the charge-changing interactions include both ($p$,$n$) and $\beta$ decay, with the latter increasing in importance with increasing $A$.}

At its most neutron-rich extent, the subsequent pathway runs parallel to and roughly $\sim 8$ neutrons away from the valley of stability for $Z>40$, with kinks at the neutron closed shells $N=82$ and $N=126$. Unlike the Wanajo $s/k>100$ cases, here most of the pathway is shifted neutron rich, even for the lighter species. The resulting abundance pattern is largely featureless, with small excesses at $A\sim 135$ and $A\sim 200$ associated with the neutron closed shells and with the original $\nu p$ pattern erased.

\begin{figure}[!htb]
	\centering
	\vspace{-5mm}
\includegraphics[width=8.5cm]{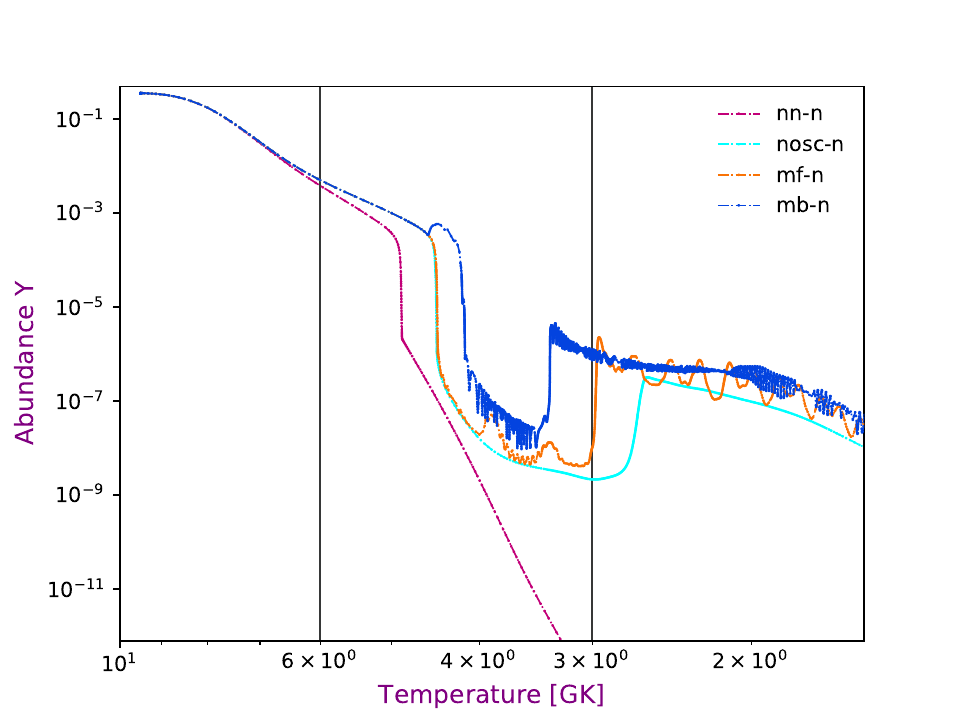}
\includegraphics[width=8.5cm]{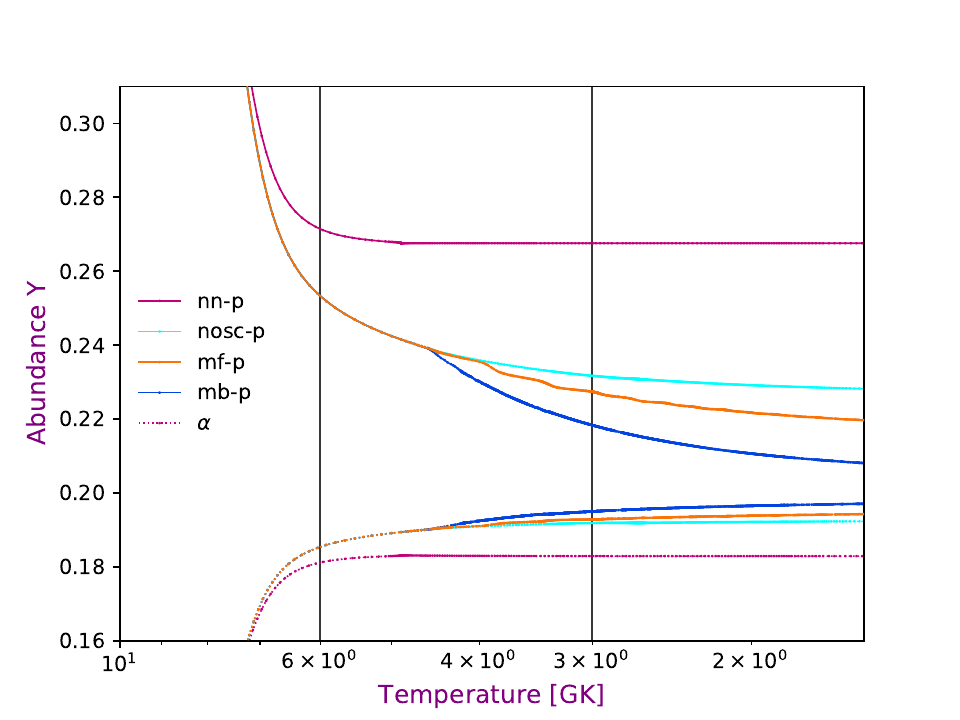}
	\caption{\it The plots of neutron (left) and proton and alpha particle (right) abundances versus temperature for the Duan2011 case with symmetric neutrino calculations (line colors as in Fig.~\ref{fig:duan2011_sym}). 
	}
	\label{fig:duan2011_sym_evol}
\end{figure}

Figure~\ref{fig:duan2011_sym_evol} shows the abundances of neutrons, protons, and alpha particles for the sym-nn, sym-nosc, sym-mf, and sym-mb calculations with the Duan2011 trajectory. Comparing to Fig.~\ref{fig:wanajo2010_sym_evol}, we note that the influence of the neutrino oscillations sets in at a later stage in the nucleosynthesis, owing to the fast expansion of the Duan2011 trajectory. Still, neutrino oscillations have greater leverage to impact the nucleosynthetic outcome here, as fewer alphas are assembled into seeds, leading to a larger free nucleon-to-seed ratio and more robust free nucleon capture. Neutron capture is in fact so robust that the transition of the nucleosynthetic pathway from proton-rich to neutron-rich occurs at a higher temperature, when photon-induced reactions are still prominent. The rise in neutron abundance after its initial dip is the establishment of an $(n,\gamma)$-$(\gamma,n)$ equilibrium on the neutron-rich side of stability. As a result, the nucleosynthetic pathway depicted in the third panel of Figure~\ref{fig:duan2011_sym_mb} shows some similarities to the shape of an $r$-process path---for example, `kinks' in the path at the $N=82$ and $N=126$ closed shells---though these nuclear features are encountered much closer to stability than in an $r$ process. {Notably, we find that the nuclear flow does not proceed past the $N=126$ shell closure for any of the $\nu i$ processes considered in this work, including those which adopt the inverted ordering for the oscillation calculations. Indeed, for the Duan2011 trajectory with higher entropy, the nucleosynthesis yields with symmetric neutrino oscillations calculations adopting inverted mass ordering are similar to the yields with normal-ordering sym-mb case.}

\subsubsection{$\nu p$- to $\nu i$-process trends}
\label{sec:vi}

\begin{figure}[!htb]
	\centering
	\vspace{-5mm}
\includegraphics[width=12cm]{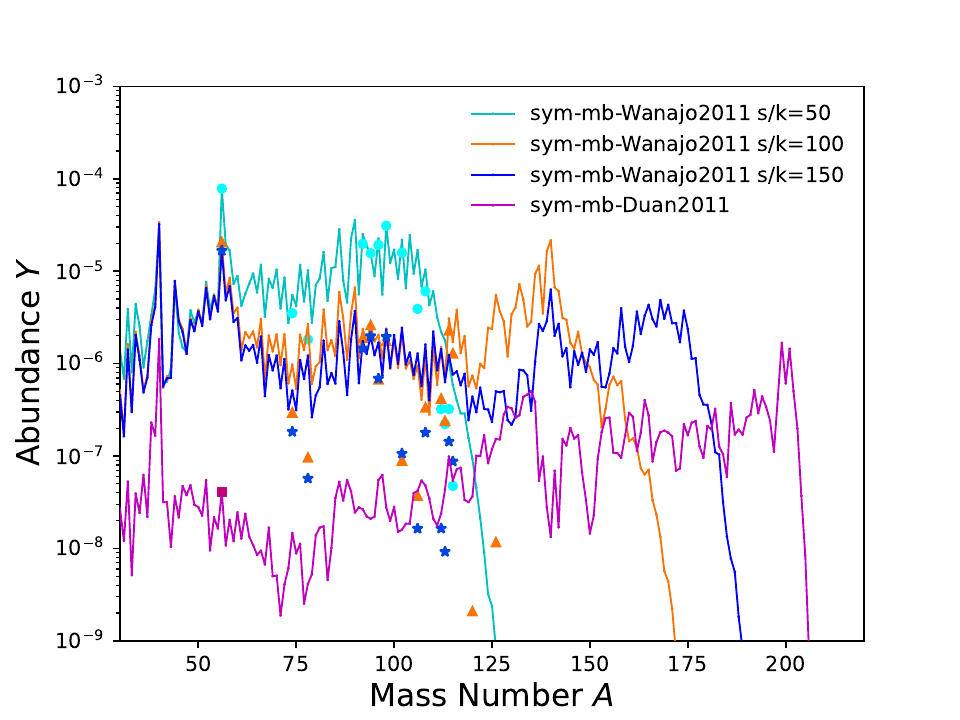}

\caption{The final abundance patterns for sym-mb calculations using Wanajo2011 $s/k=50$ (light blue line), 100 (orange line), 150 (blue line) and Duan2011 (purple line) trajectories. The colored data points indicate the abundances of the proton-rich $p$ nuclei synthesized in each case, respectively.
}
	\label{fig:ab_sym_wanajo_duan_pn}
\end{figure}

As discussed above, we find that the influence of neutrino oscillations grows with the increase of initial entropy, and the ultimate extent and yields of the element synthesis vary from a typical $\nu p$ process to an unusual mixture of light proton-rich nuclei and heavy neutron-rich nuclei to a full and robust $\nu i$ process.  To examine this trend, Fig.~\ref{fig:ab_sym_wanajo_duan_pn} shows the comparison of the final abundance yields from the Wanajo2011 trajectories with varying entropy as well as the Duan2011 trajectory, adopting the symmetric many-body (sym-mb) neutrino calculations that result in the greatest influence of the neutrino interactions.  Also shown (as points) are the abundances of the proton-rich $p$ nuclei synthesized in each calculation, to highlight the fraction of the material that ultimately shifts neutron-rich. 
Note the abundance pattern of Wanajo2011 $s/k =50$ case follows a typical $\nu p$ process where $p$ nuclei are dominantly produced, while the abundance patterns resulting from larger initial entropy values shift from a $\nu p$ process at lower mass to a neutron-rich pattern for heavier nuclei ($A\gtrsim 115$ for $s/k =100$, $A\gtrsim 100$ for $s/k =150$, $A\gtrsim 70$ for Duan2011).

\begin{figure}[!htb]
	\centering
	\vspace{-5mm}
\includegraphics[width=12cm]{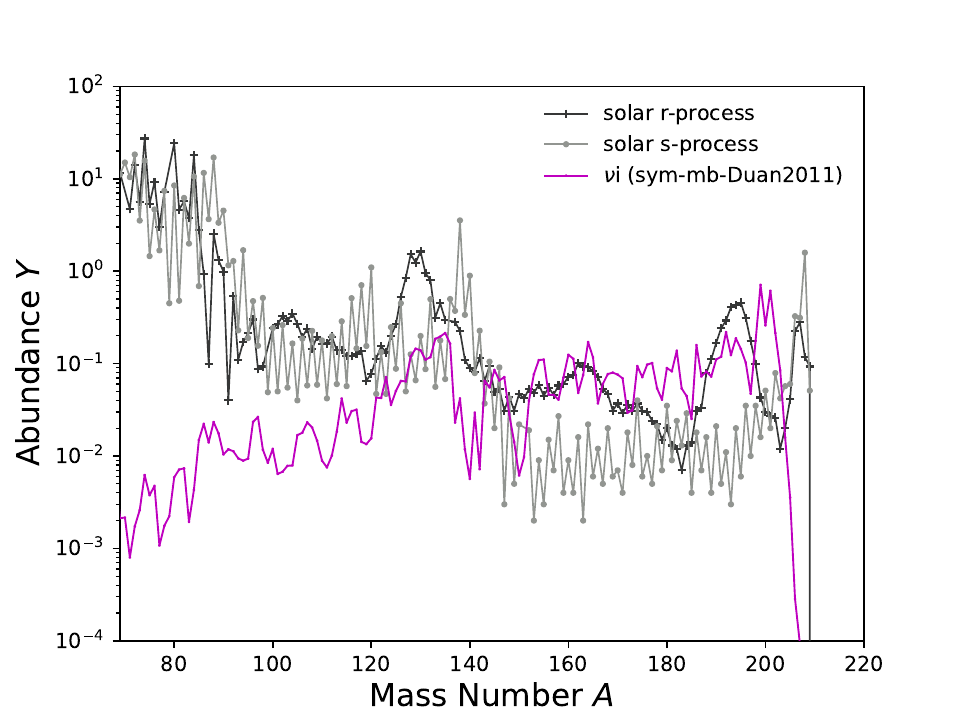}
\caption{The final abundance pattern for sym-mb calculation with Duan2011 ($\nu i$-process pattern; purple line), compared with the solar system $s$-process (grey line) and $r$-process (black line) abundance data~\citep{Sneden2008}. The $\nu i$ abundance for $A=143$ is scaled to the solar $r$-process data for pattern comparison.}
	\label{fig:ab_duan_solar}
\end{figure}

To further explore the features of these novel $\nu i$-process patterns, we also compare the resulting abundance patterns of the simulations with the Duan2011 matter trajectory and sym-mb neutrino calculations to the solar system 
separated $s$-process and $r$-process patterns~\citep{Sneden2008} in Fig.~\ref{fig:ab_duan_solar}. 
Given that its nucleosynthetic pathway is between those of the $s$ process and $r$ process, the $\nu i$ process abundances are distinct from those of both the solar $s$ process and $r$ process, showing shifted neutron closed shell features and a distinctly higher lanthanide production than the $s$ process.

In summary, from the nucleosynthesis tests of the supernova neutrino-driven wind trajectories with {various neutrino treatments}, using a neutrino energy distribution symmetric in $\nu_e$ and $\bar\nu_e$ (to ensure proton-rich conditions), we see the emergence of a novel type of heavy-element nucleosynthesis: a $\nu i$ process. From this first exploratory study, we conclude that a $\nu i$ process can occur in high-entropy ($s/k\gtrsim$100), neutrino-rich environments with initially proton-rich conditions. This process results in abundance yields that can be a mixture of a $\nu p$-process-like pattern at lower mass and an $i$-process-like pattern for heavier elements, or an entirely $i$-process-like pattern for all elements. Since the transition to neutron-capture nucleosynthesis results from the continued production of neutrons by neutrino reactions, this process is highly sensitive to the neutrino oscillation physics.

\subsection{Heavy-element nucleosynthesis with asymmetric neutrino calculations}
\label{sec:asymmetric}

The asymmetric neutrino calculations adopted in Table~\ref{tab:neutrinomodels} result in capture rates for antineutrinos that are similar to or higher than for neutrinos, producing equilibrium electron fractions that are closer to or less than 0.5. As already noted, the supernova neutrino-driven wind could be a mildly neutron-rich environment that allows for a limited or weak $r$ process, with the nucleosynthesis extent sensitive to the neutrino physics adopted. So, here we investigate a possible $r$ process in the neutrino-driven wind and the effect of the neutrino physics treatments described above on the $r$-process yields.  
We examine the nucleosynthetic impact of neutrinos with asymmetric energy distributions and luminosities with four different choices of neutrino energy distributions, i.e.,  asym2, asym2.1, asym3, and asym4, corresponding to equilibrium $Y_e$ values ranging from 0.504 to 0.366, as listed in Table~\ref{tab:neutrinomodels}.

Three of the four adopted sets of asymmetric neutrino energies result in initially neutron-rich conditions. During alpha particle formation, all of the protons are consumed to produce alpha particles. This reaction leaves few free protons to convert to neutrons via antineutrino capture, while any excess free neutrons are still subject to neutrino captures. The protons produced as a result combine promptly with free neutrons to continue forming alphas. Thus neutrino interactions act to increase the number of alpha particles (and, subsequently, seed nuclei) and reduce the number of free neutrons available for capture on the seed nuclei, which is the so-called alpha effect \citep{Fuller+1995, McLaughlin+1996, Meyer1998}. With the influence of collective neutrino oscillations included, the effective energies of $\nu_e$ and $\bar{\nu}_e$ are raised. The corresponding increases to the neutrino capture rates cause the alpha effect to become more significant and the production of heavy nuclei is hindered.

\begin{figure}[!htb]
	\centering
	\vspace{-5mm}
\includegraphics[width=7.5cm]{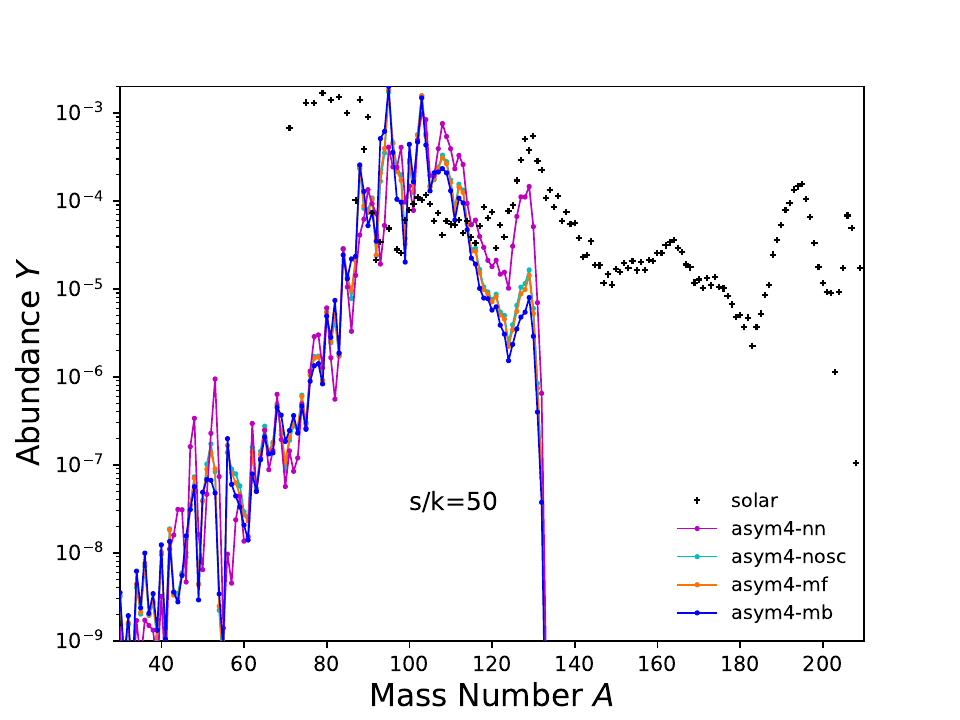}
\includegraphics[width=10cm]{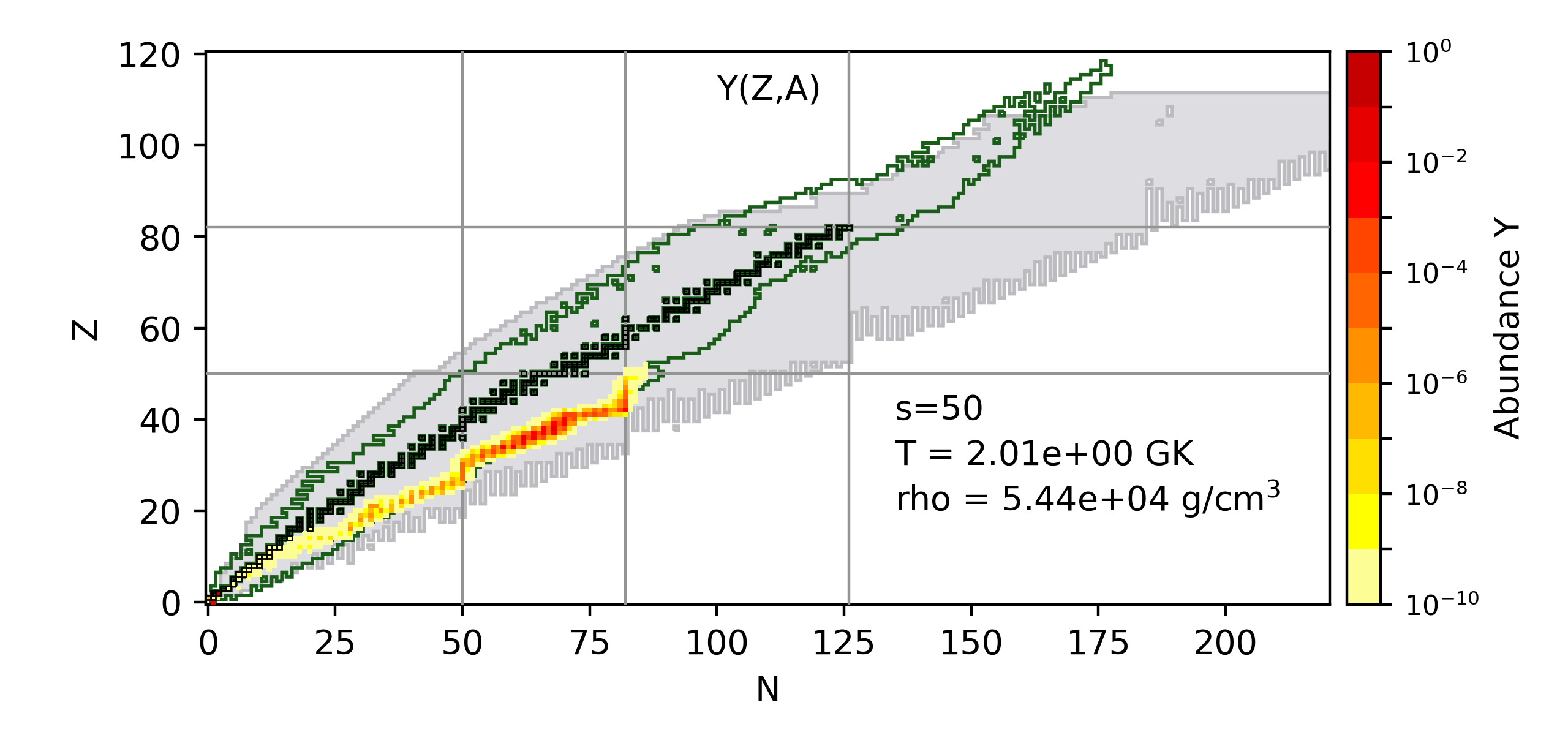}\\
\includegraphics[width=7.5cm]{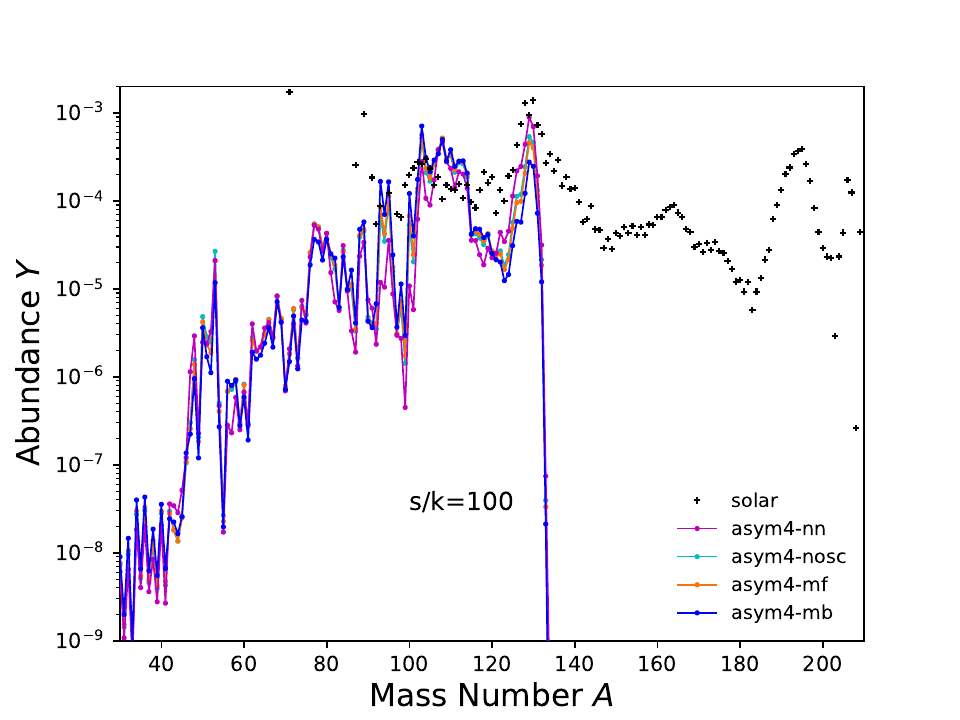}
\includegraphics[width=10cm]{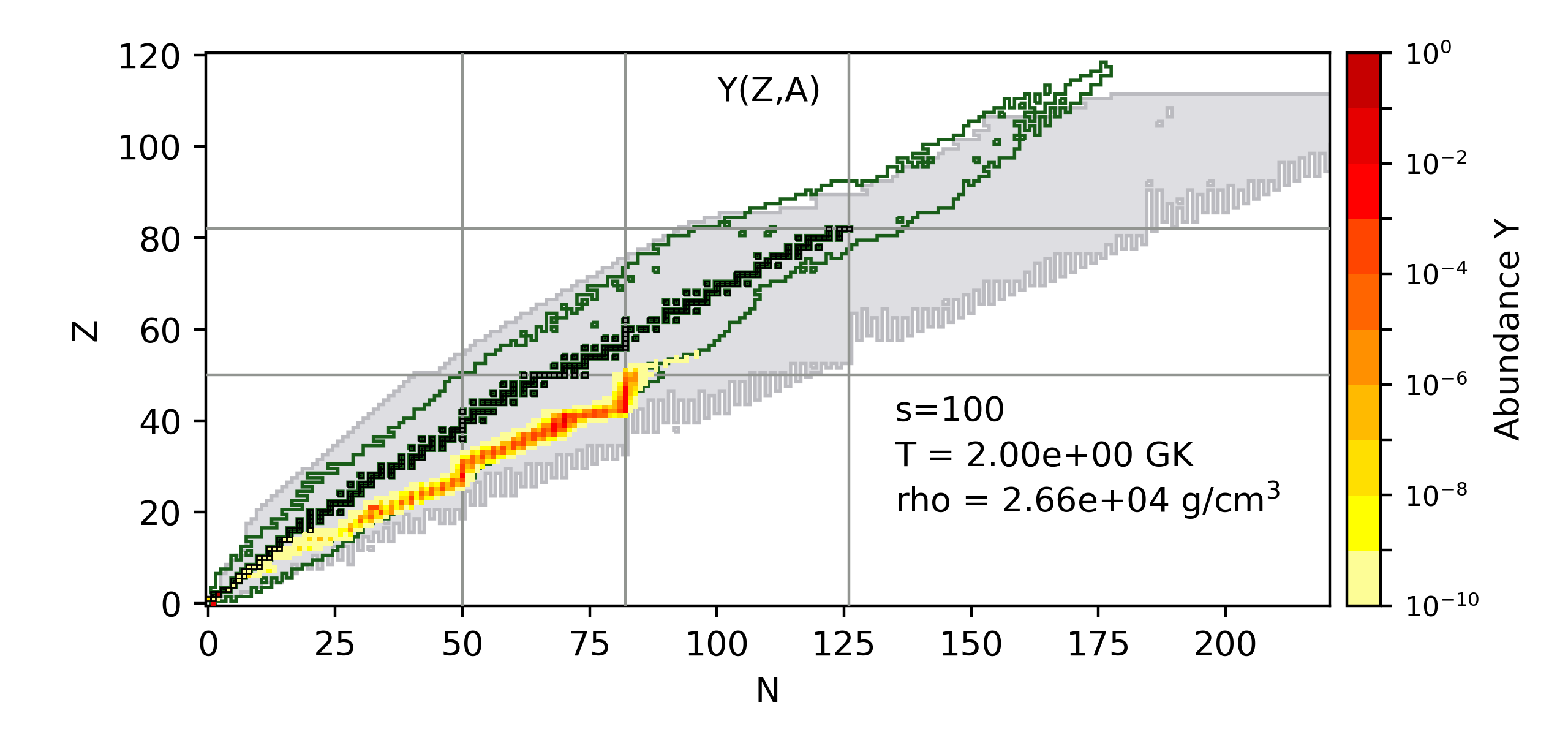}\\
\includegraphics[width=7.5cm]{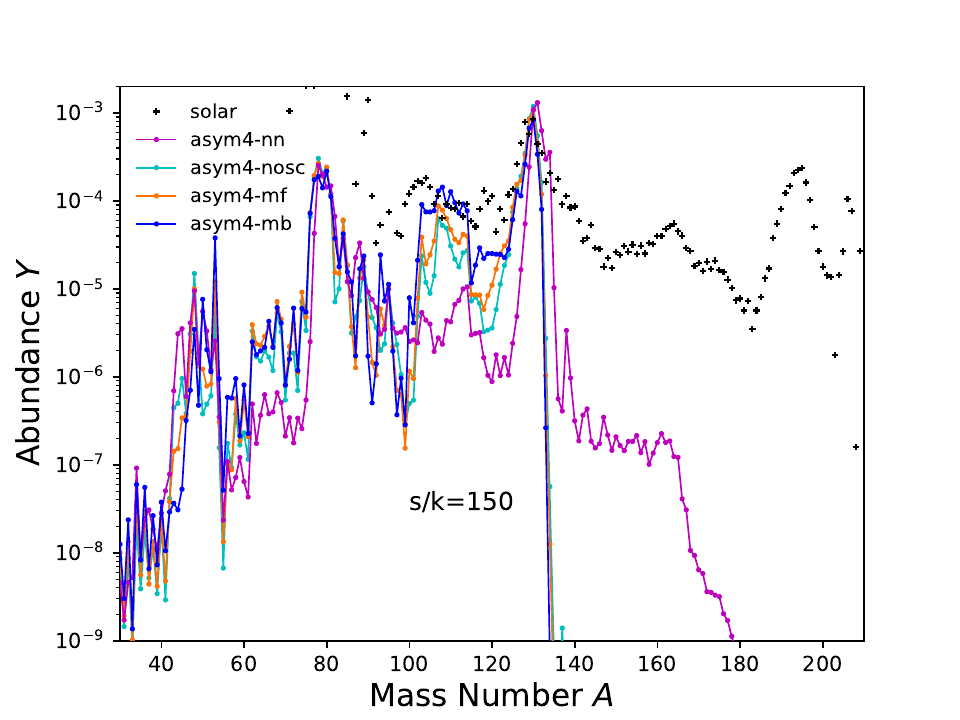}
\includegraphics[width=10cm]{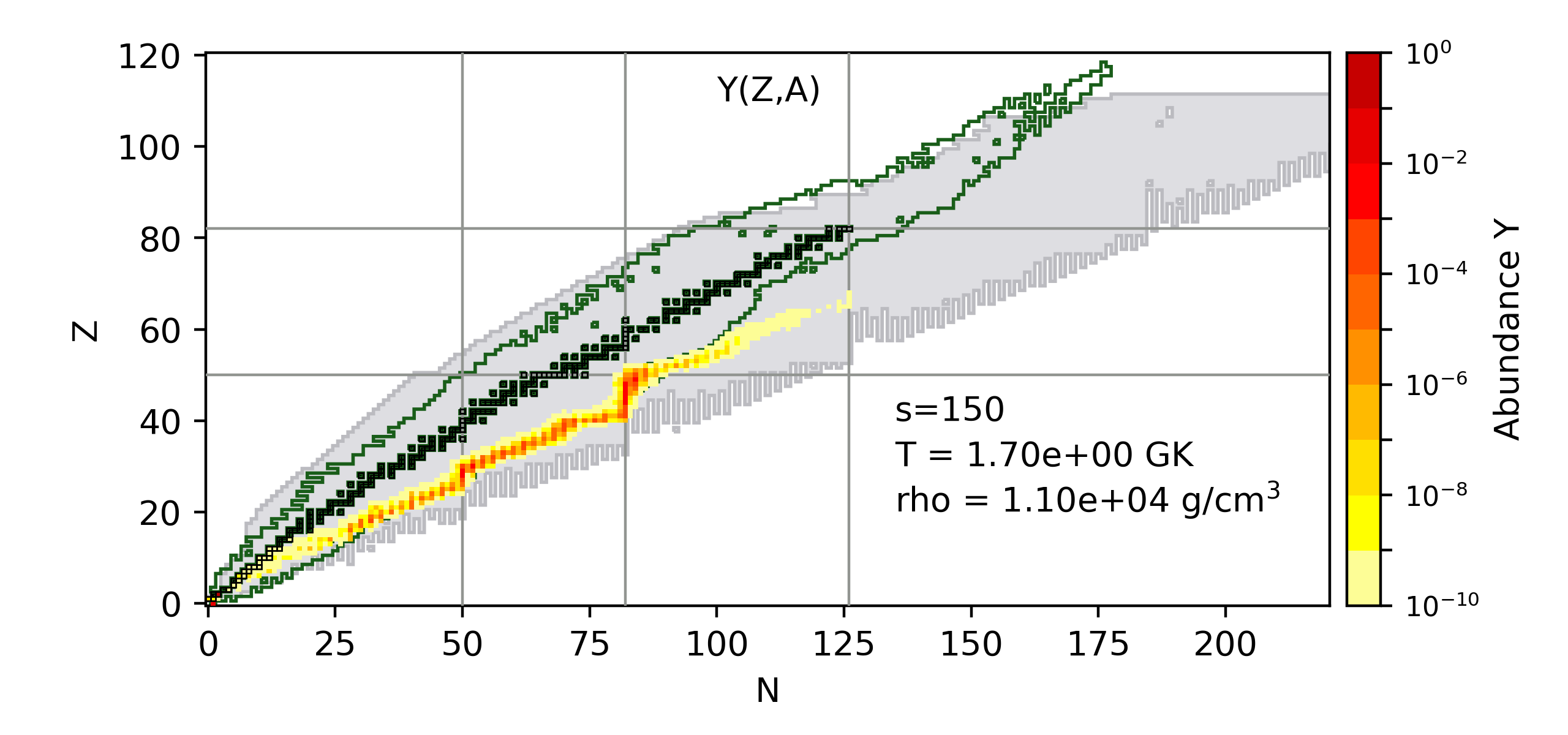}\\
	\caption{The abundance patterns plotted as functions of the atomic weight $A$ (left) and nucleosynthesis path when is farthest from stability (right), for simulations with the Wanajo2011 trajectories with asymmetric neutrino calculations asym4 (purple: nn for no neutrinos involved for T $<$ 10 GK; cyan: nosc for no neutrino oscillations; orange: mf for mean-field neutrino oscillations; blue: mb for many-body neutrino oscillations), for varying entropy. 
	\label{fig:wanajo10_asym}
    }
\end{figure}

We first test the nucleosynthesis calculations with the asymmetric neutrino treatment asym4, which results in the most neutron-rich conditions of the four cases and therefore is most likely to produce $r$-process species. The corresponding abundance patterns and the nucleosynthesis paths are shown in Fig.~\ref{fig:wanajo10_asym}, for the $s/k=50$, 100, and 150 Wanajo11 matter trajectories. All simulations result in at most a weak $r$ process, and only the $s/k=150$ case robustly produces the second ($A\sim130$) $r$-process peak. The influence of neutrinos and their oscillations only act to hinder the $r$ process through an enhanced alpha effect. The asym4-mb case shows the largest influence, though overall the impact of neutrinos on the ultimate nucleosynthetic outcome is much smaller compared to the simulations discussed in Sec.~\ref{sec:sym}. The free nucleon-to-seed ratio is low and the neutrons are efficiently consumed, cutting off the potential influence of neutrinos. This situation is in stark contrast to the proton-rich examples above in which protons are available to convert to neutrons throughout the element synthesis.

\begin{figure}[!htb]
	\centering
	\vspace{-5mm}
\includegraphics[width=8cm]{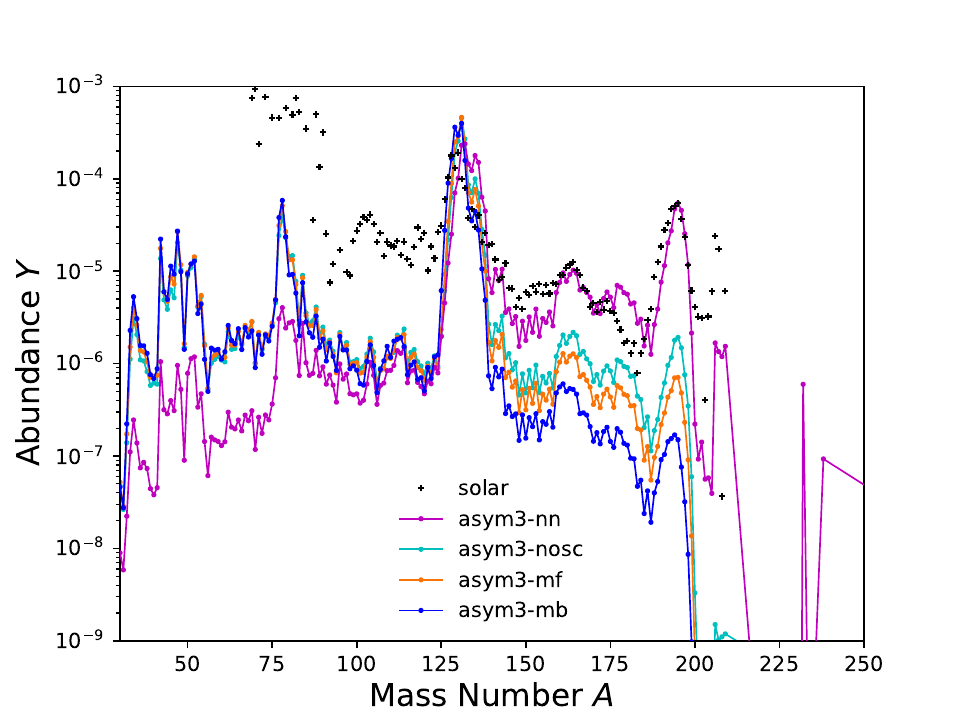}
\includegraphics[width=8cm]{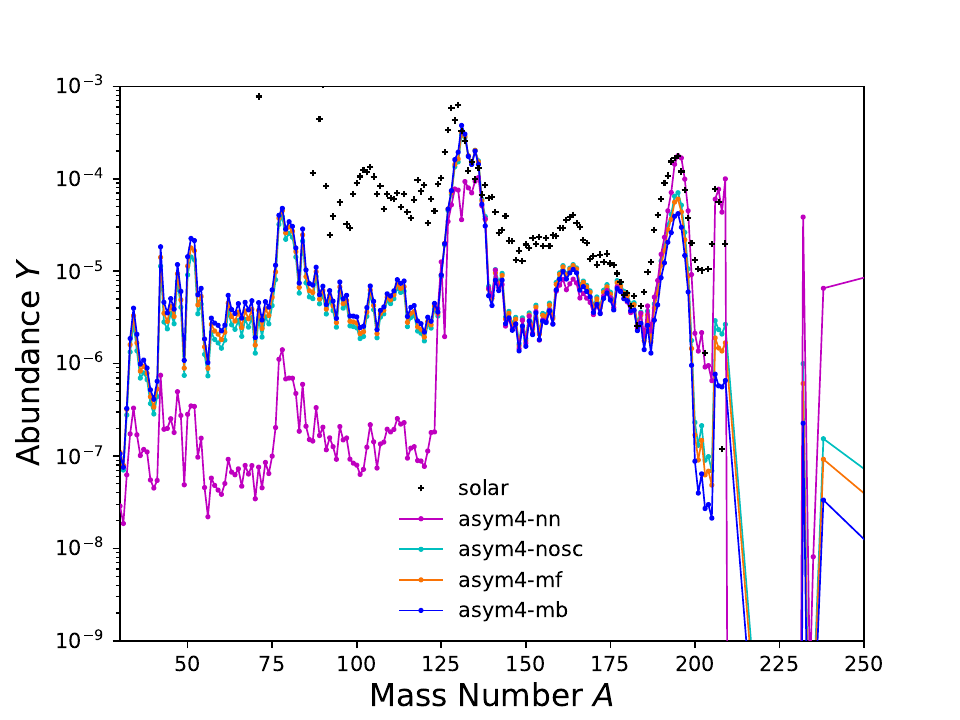}\\
	\caption{The abundance patterns of simulations with the Duan2011 matter trajectory and asymmetric neutrino calculations asym3 (left) and asym4 (right) plotted as functions of the atomic weight $A$ (line colors as in Fig.~\ref{fig:wanajo10_asym}). 
	}
	\label{fig:duan2011_asym2}
\end{figure}

In order to fully explore the potential impact of the different neutrino treatments (nn, nosc, mf, mb) defined in Sec.~\ref{sec:nucalc}, we also calculate the nucleosynthesis resulting from the Duan2011 matter trajectory. Its fast expansion and high entropy ensure that a robust main $r$ process---one that reaches $A\sim195$ peak nuclei---is possible for at least some of the neutrino energy distributions explored, and the impact of the neutrino treatments is maximized for all asymmetric configurations explored: asym4, asym3, asym2.1, and asym2.

\begin{figure}[!htb]
	\centering
	\vspace{-5mm}
\includegraphics[width=8.5cm]{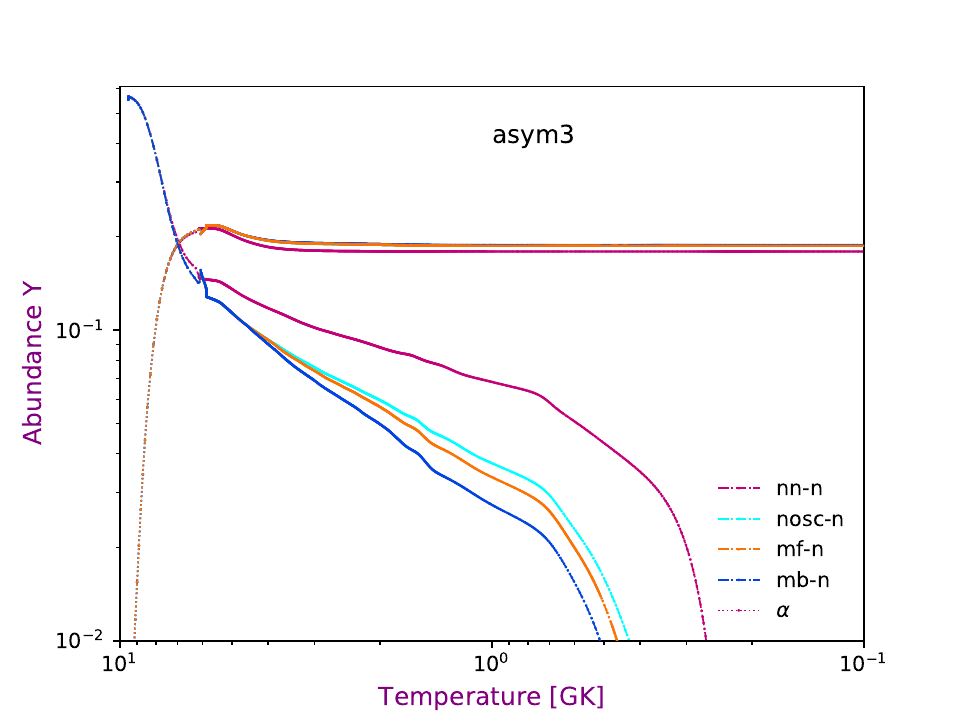}
\includegraphics[width=8.5cm]{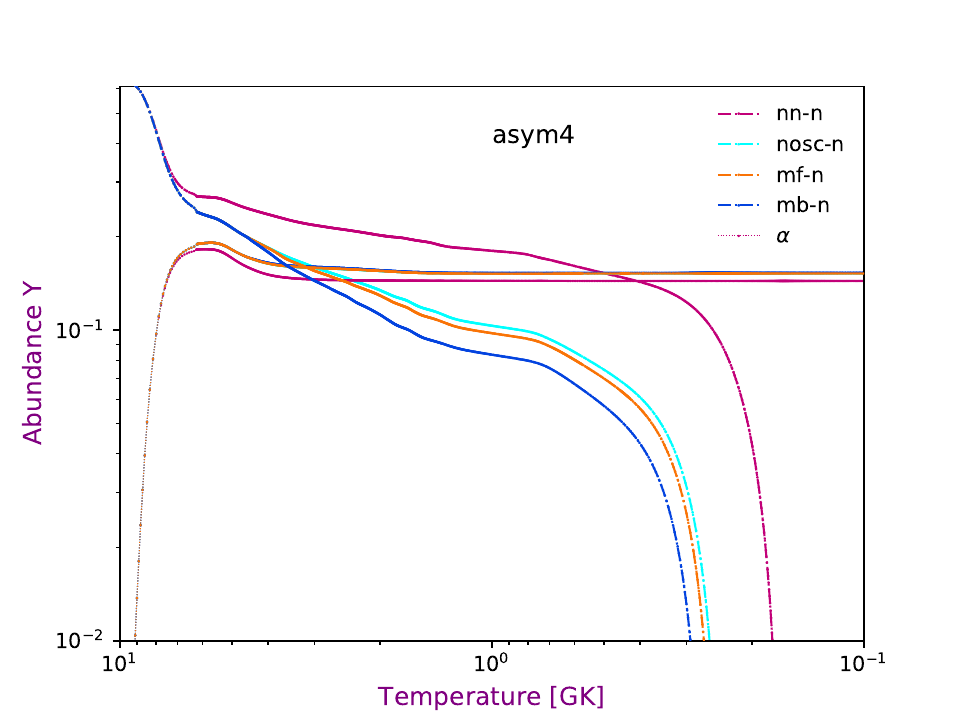}
\\
	\caption{Abundances of free neutrons and alpha particles versus temperature with neutrino cases asym3 (left) and asym4 (right) with the Duan2011 matter trajectory (line colors as in Fig.~\ref{fig:wanajo10_asym}).} 
	\label{fig:duan2011_asym_abn}
\end{figure}

For the most neutron-rich asym4 case, as seen in the right panel of Fig.~\ref{fig:duan2011_asym2}, the nucleosynthesis calculation for the Duan2011 matter trajectory with neutrino interactions turned off for $T<10$ GK results in a robust main $r$-process pattern that reaches the actinide region. Neutrino interactions produce an alpha effect of varying strength, as shown in Fig.~\ref{fig:duan2011_asym_abn}, with the asym4-mb case resulting in the lowest free neutron abundance and highest alpha abundance. As a result, we observe decreased production of heavier nuclei especially around the third-peak region and beyond, with the smallest $A\sim 195$ peak produced in the asym4-mb case. 

The asym3 case starts with a slightly higher $Y_e$ value at equilibrium, producing a lower initial neutron abundance and a faster consumption of neutrons than in the asym4 case, as shown in Figure~\ref{fig:duan2011_asym_abn}. As a result, the abundance pattern for the asym3-nn case just barely fills out the third-peak region and underproduces the actinides. Again, from the left panel of Fig.~\ref{fig:duan2011_asym2}, we can see that neutrinos and neutrino oscillations hinder the $r$-process production of heavier nuclei. The effect of the difference in neutrino treatments on the $r$-process yields follows the same pattern as the asym4 case, but with more significant deviations shown.

\begin{figure}[!htb]
	\centering
	\vspace{-5mm}
\includegraphics[width=8cm]{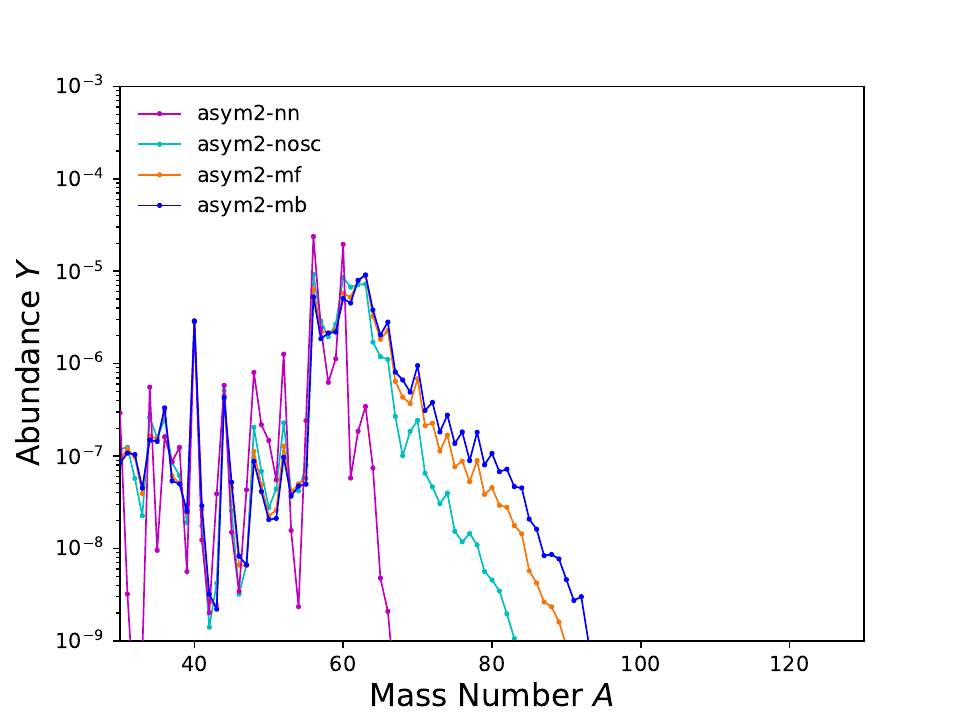}
\includegraphics[width=8cm]{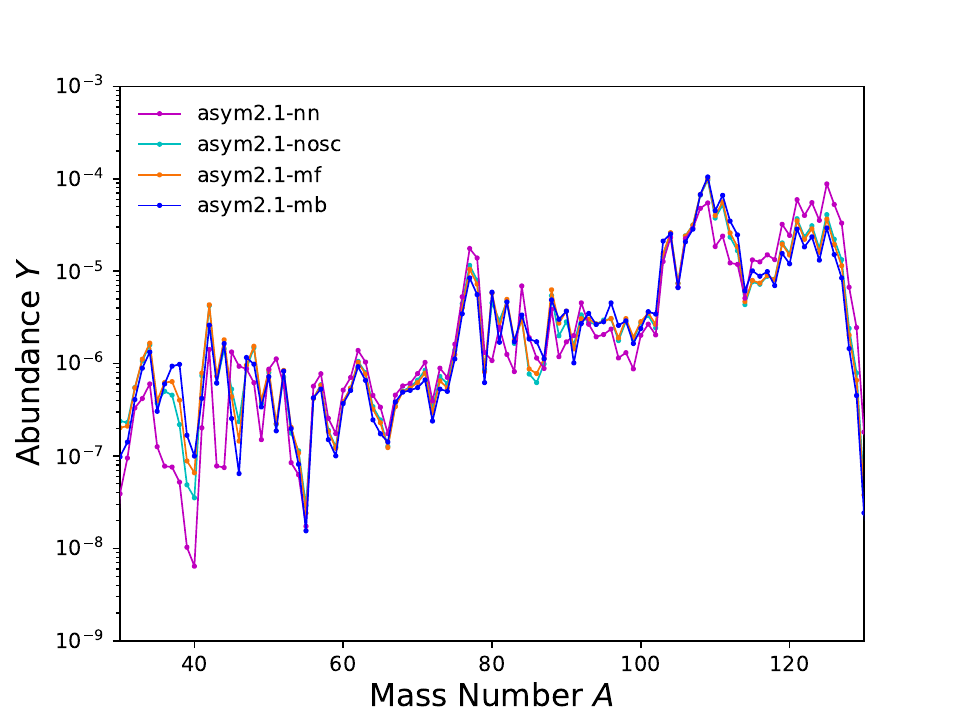}\\
	\caption{The abundance patterns of simulations with the Duan2011 trajectory with asymmetric neutrino calculations asym2 (left) and asym2.1 (right), plotted as functions of the atomic weight $A$ (line colors as in Fig.~\ref{fig:wanajo10_asym}). 
	}
	\label{fig:duan2011_asym1}
\end{figure}

Finally, we examine the effects of the different neutrino treatments on the nucleosynthesis calculations for the Duan2011 matter trajectory with initial $Y_e$ values around 0.5, with the results shown in Fig.~\ref{fig:duan2011_asym1}. For the neutrino parameters that result in an electron fraction slightly bigger than 0.5, the nucleosynthesis result is either iron-peak nuclei (for asym2-nn), or a mild $\nu p$ process when neutrino interactions are included. Similar to the symmetric neutrino cases described in Section~\ref{sec:vpprocess} above, the extent of the $\nu p$ process is sensitive to the treatment of neutrino oscillations, with the asym2-mb case producing the highest $\nu p$-process yields compared to the other neutrino treatments. The asym2.1 neutrino distribution on the other hand corresponds to an equilibrium electron fraction slightly smaller than 0.5 and result in a weak $r$ process up to the second $r$-process peak. Here the influence of neutrinos is modest, similar to the Wanajo2011 asym4 results shown in Fig.~\ref{fig:wanajo10_asym}.

For the asymmetric neutrino conditions studied here, we find the neutrino physics to be particularly crucial for the nucleosynthesis outcomes that sit at a threshold, either for an $r$ process that barely reaches the third peak region, or for initial conditions around $Y_e=0.5$. We accordingly expect similar behavior for an $r$ process that proceeds just beyond the second peak. In these cases, the changes to the neutrino interaction rates that result from the treatment of neutrino oscillation behavior can produce marked changes in the abundance yields. However, in general, the leverage that the neutrinos have on nucleosynthetic outcomes in neutron-rich conditions remains smaller than in very proton-rich conditions. This difference is due to neutrino interactions changing the abundance of the sub-dominant free nucleon species, e.g., neutrons in proton-rich and protons in neutron-rich nucleosynthesis, as element synthesis proceeds. However, as the temperature drops, only neutrons can be captured (and protons cannot). Thus, the influence of neutrino interactions in proton-rich conditions is extended throughout the element synthesis, resulting in a larger impact on a $\nu p$ process compared to the $r$ process for the collective oscillation effects considered here.

\section{Discussion and Conclusions}

Collective neutrino oscillations, driven by neutrino-neutrino interactions, can lead to substantial flavor conversion near the supernova core, with potentially significant implications for nucleosynthesis as well as energy transport within the supernova envelope~\cite{Duan2009, Duan2010, Chakraborty+2016, Tamborra2020, Richers2022, Wang2023,Volpe:2023met, Patwardhan:2021rej, Balantekin:2023qvm}.
In this work, we explored the impact of two different treatments of collective neutrino oscillations (many-body and mean-field), along with two other treatments (no neutrino interactions, and neutrino interactions without oscillations) as comparison baselines, on the heavy-element nucleosynthesis in a supernova neutrino-driven wind, in both proton-rich and neutron-rich conditions. 
We adopted a simplified interacting neutrino system to examine a many-body calculation as well as the corresponding mean-field approximation of the collective oscillation effects. 
To examine different nucleosynthesis regimes, we varied the neutrino energy distributions to generate a range of initial neutron-to-proton ratios, and, for each set of neutrino energy distributions, we ran nucleosynthesis calculations for the four different neutrino treatments mentioned above. 

We found that the difference in the neutrino treatments has the largest impact on \textit{proton-rich nucleosynthesis}, particularly at high entropies. In our setup, this corresponded to a symmetric choice of neutrino vs antineutrino energies. Here, neutrino oscillations facilitate the synthesis of increasingly heavy elements.
Indeed, for the supernova trajectories with entropy values of  $s/k = 100$ or above, the neutrinos can promote further synthesis of heavier elements via late-time neutron capture through a novel $\nu i$ process, particularly when neutrino oscillations are included. The $\nu i$ process can result in a mixture of a $\nu p$-process-type pattern at
lower mass and an $i$-process-like pattern at higher mass, or a fully $i$-process-like pattern at the highest entropies. This process is very sensitive to the neutrino oscillation treatment, and we found that the many-body treatment of collective oscillations resulted in maximum enhancement of element synthesis among the treatments explored. We plan to explore the implications of a potential $\nu i$ process in future work.

For asymmetric neutrino and antineutrino energy configurations, we found the effects of different neutrino treatments on the nucleosynthesis yields to be somewhat smaller. However, the neutrino physics is crucial particularly for conditions near a threshold, e.g., when $Y_e$ is around 0.5, or when the conditions are such that the third $r$-process peak is just barely reached. Similarly to the proton-rich case, the many-body calculation of collective oscillations results in the largest difference in the $r$-process nucleosynthesis yields compared to the calculation without neutrinos, by hindering the production of heavy elements through an enhanced alpha effect.

In all cases examined here, different treatments of collective neutrino oscillations have a non-negligible impact on the heavy-element nucleosynthesis yields in supernovae. In addition to the collective neutrino oscillations, other types of flavor transformation could also occur and may influence nucleosynthesis, e.g., active-sterile conversions \citep{McLaughlin1999, Fetter:2002xx, Wu2014, Pllumbi2015}, or fast neutrino flavor conversion \citep{Wu2014,Xiong:2020ntn,Fujimoto:2022njj}. Moreover, other physical phenomena besides neutrino flavor transformations could also have significant impacts on nucleosynthesis---for instance, neutrino interactions induced by magnetic moments (\cite{Balantekin:2007xq}), or uncertainties in rotation rate and magnetic fields~\citep{Vincenzo:2021rvw} or nuclear reaction rates~\citep{Psaltis:2022jgr} in neutrino-driven wind environments. A better understanding and implementation of the neutrino physics in candidate heavy-element nucleosynthesis events is necessary and crucial to solve the puzzle of the origins of the heavy elements.

\section{Acknowledgements}
The authors would like to thank I.~Roederer and A.~Friedland for helpful discussions. This research is supported in part by the National Science Foundation Grants No. PHY-1630782 and PHY-2020275 (Network for Neutrinos, Nuclear Astrophysics and Symmetries).
ABB is supported in part by the U.S.~Department of Energy, Office of Science, Office of High Energy Physics, under Award  No.~DE-SC0019465 and in part by the National Science Foundation Grant PHY-2108339 at the University of Wisconsin-Madison. 
The work of M.J.C.~is supported by the U.S.~Department of Energy under contract number DE-SC0021143. 
The work of AVP was supported by the U.S. Department of Energy under contract numbers DE-AC02-76SF00515 at SLAC National Accelerator Laboratory and DE-FG02-87ER40328 at the University of Minnesota. 
The work of R.S is supported in part by the U.S. Department of Energy under contract numbers DE-FG02-95-ER40934 and LA22-ML-DE-FOA-2440. 
The work of X.W. is supported in part by the Chinese Academy of Sciences (Grant No. E329A6M1) and the National Natural Science Foundation of China (Grant No. E311445C). 
We would like to acknowledge the Mainz Institute for Theoretical Physics (MITP) of the Cluster of Excellence PRISMA+ (Project ID 39083149), for enabling us to complete this work. This research was also supported in part by the National Science Foundation under Grant No.PHY-1748958 at the Kavli Institute for Theoretical Physics (KITP), and by the U.S. Department of Energy grant No. DE-FG02-00ER41132 at the Institute for Nuclear Theory (INT). The authors would like to thank KITP, MITP, and INT for their hospitality and support at different stages during the completion of this work.

\bibliography{neutrino}
\end{document}